\documentstyle{mn}
\input psfig.sty
\def\etal{\it et al. \rm}
\def\kms{km~s$^{-1} \,$}

\def\hmpc{{\rm h}$^{-1}~{\rm Mpc}\,$}
\def\hkpc{{\rm h}$^{-1}~{\rm kpc}\,$}
\def\Mo{{\rm M_\odot}}

\def\gs{\mathrel{\raise0.35ex\hbox{$\scriptstyle >$}\kern-0.6em 
\lower0.40ex\hbox{{$\scriptstyle \sim$}}}}
\def\ls{\mathrel{\raise0.35ex\hbox{$\scriptstyle <$}\kern-0.6em 
\lower0.40ex\hbox{{$\scriptstyle \sim$}}}}
\def\gsim{ \lower .75ex \hbox{$\sim$} \llap{\raise .27ex \hbox{$>$}} }
\def\lsim{ \lower .75ex \hbox{$\sim$} \llap{\raise .27ex \hbox{$<$}} }
\def\LCDM{\hbox{$\Lambda$CDM }}
\def\P3M{\hbox{$P^{3}M$}}

\def\AP3M{\hbox{$AdP^{3}M$}}

\def\cc2{c2}
\def\cc3{c3}
\def\cc4{c4}
\def\cc{c}

\def\aap{A\&A}

\def\AJ{AJ}

\def\ApJ{ApJ}
\def\apj{ApJ}
\def\ApJL{ApJL}
\def\apjl{ApJL}
\def\ApJS{ApJS}

\def\MN{MNRAS}
\def\mnras{MNRAS}

\def\etal{{\it et al.\thinspace}}
\def\eg{{\it e.g.\ }}

\def\spose#1{\hbox to 0pt{#1\hss}}
\def\approxlt{\mathrel{\spose{\lower 3pt\hbox{$\sim$}}
	\raise 2.0pt\hbox{$<$}}}
\def\approxgt{\mathrel{\spose{\lower 3pt\hbox{$\sim$}}
	\raise 2.0pt\hbox{$>$}}}

\def\<{\thinspace}

\def\boxit#1{\vbox{\hrule\hbox{\vrule\kern3pt\vbox{\kern3pt
          #1 \kern3pt}\kern3pt\vrule}\hrule}}

\def\K{{\rm\thinspace K}}

\def\kpc{{\rm\thinspace kpc}}

\def\Mpc{{\rm\thinspace Mpc}}
\def\hmpc{{\rm\thinspace $h^{-1}\rm\thinspace Mpc$\ }}
\def\hkpc{{\rm\thinspace $h^{-1}\rm\thinspace kpc$\ }}
\def\Msun{\hbox{$\rm\thinspace M_{\odot}$}}

\def\h50{\hbox{$\rm\thinspace h_{50}$}}
\def\h50m1{\hbox{$\rm\thinspace h_{50}^{-1}$}}

\date{Revised Version}
\pagerange{\pageref{firstpage}--\pageref{lastpage}}
\pubyear{2000}
\begin{document}
\label{firstpage}
\title[Cosmological galaxy formation]
{Simulations of galaxy formation in a cosmological volume}

\author[F. R. Pearce \etal]
{F. R. Pearce$^{1,7,8}$, A. Jenkins$^1$, C. S. Frenk$^1$, S. D. M. White$^2$, P. A. Thomas$^3$, \\
\newauthor H. M. P. Couchman$^4$, J. A. Peacock$^5$, G. Efstathiou$^6$ (The Virgo Consortium) \\
$^1$ Physics Department, South Rd, Durham, DH1 3LE, UK \\
$^2$ Max-Plank-Institut fur Astrophysik, Garching, Germany \\
$^3$ Astronomy Centre, CPES, University of Sussex, Falmer, Brighton, BN1 9QJ, UK \\
$^4$ Department of Physics and Astronomy, McMaster University, Hamilton, Ontario, L8S 4M1, Canada \\
$^5$ Institute for Astronomy, University of Edinburgh, Edinburgh, EH9 3HJ, UK \\
$^6$ Institute of Astronomy, Madingley Rd, Cambridge, UK \\
$^7$ Physics and Astronomy, University of Nottingham, Nottingham, UK \\
$^8$ email:F.R.Pearce@durham.ac.uk \\
}

\maketitle

\begin{abstract} 

We present results of large $N-$body-hydrodynamic simulations of galaxy
formation. Our simulations follow the formation of galaxies in cubic
volumes of side $100\Mpc$, in two versions of the cold dark matter (CDM)
cosmogony: the standard, $\Omega=1$ SCDM model and the flat, $\Omega=0.3$
\LCDM model. Over 2000 galaxies form in each of these simulations. We
examine the rate at which gas cools and condenses into dark matter
halos. This roughly tracks the cosmic star formation rate inferred from
observations at various redshifts. Galaxies in the simulations form
gradually over time in the hierarchical fashion characteristic of the CDM
cosmogony.  In the \LCDM model, substantial galaxies first appear at
$z\simeq 5$ and the population builds up rapidly until $z=1$ after which
the rate of galaxy formation declines as cold gas is consumed and the
cooling time of hot gas increases. In the SCDM simulation, the evolution is
qualitatively similar, but it is shifted towards lower redshift. 
In both cosmologies, the present-day K-band luminosity function of the
simulated galaxies resembles observations. The galaxy autocorrelation
functions differ significantly from those of the dark matter. At the
present epoch there is little bias in either model between galaxies and
dark matter on large scales, but a significant anti-bias on scales of
$\sim1$\hmpc and a positive bias on scales of $\sim 100$\hkpc.  The galaxy
correlation function evolves little with redshift in the range $z=0 -3$,
and depends on the luminosity of the galaxy sample. The projected pairwise
velocity dispersion of the galaxies is much lower than that of the dark
matter on scales less than 2\hmpc. Applying a virial mass estimator to the
largest galaxy clusters recovers the cluster virial masses in an unbiased
way.  Although our simulations are affected by numerical limitations, they
illustrate the power of this approach for studying the formation of the
galaxy population.

\end{abstract} 


\section{Introduction}

A detailed understanding of galaxy formation is one of the central goals of
contemporary astrophysics. Over the past decade, this goal seems, finally,
to have come within reach. On the observational side, data from the Keck
and Hubble Space telescopes have revolutionised our view of the high
redshift Universe.  From the theoretical point of view, it is clear that
studying formation involves a synthesis of ideas from a wide range of
specialities.  A full theoretical treatment requires consideration of the
early Universe processes that created primordial density fluctuations, the
non-linear dynamics that result in the formation of dark matter halos, the
dissipational processes of cooling gas, the microphysics and chemistry that
precipitate star formation, the feedback due to energy exchange between
supernovae and interstellar gas, the merging of galaxy fragments and the
interaction of galaxies with their large-scale environment.

Because of its non-linear character, lack of symmetry and general
complexity, galaxy formation is best approached theoretically using
numerical simulations. As a minimum, realistic simulations must be able to
follow the evolution of the dark matter from appropriate initial
conditions, together with the coupled evolution of a dissipative gas
component that will eventually produce the visible regions of galaxies.
The advent of large computers, particularly parallel supercomputers,
together with the development of efficient algorithms, has given
considerable impetus to this approach in recent years.

Both Eulerian and Lagrangian gas dynamics techniques have been used to
simulate galaxy formation.  Currently, only the latter have achieved
sufficient dynamic range to resolve galaxy formation within
cosmological volumes.  For example, the recent large Eulerian
simulations of Blanton \etal (1999) have spatial resolution of order 
$300 - 500 \kpc$, whereas typical Lagrangian techniques can
readily achieve spatial resolutions of order $20\kpc$ or better
(e.g. Carlberg, Couchman \& Thomas 1990, Katz, Hernquist \& Weinberg 1992,
1999, 
Evrard, Summers \& Davis 1994, Frenk \etal 1996, Navarro \& Steinmetz
1997). Recently, the power of the traditional, fixed-grid Eulerian
approach has been greatly enhanced by the inclusion of adaptive mesh
refinement techniques which have been applied to
the study of the early phases of galaxy formation (Abel \etal 1998).

In this paper, we carry out simulations of galaxy formation using the
Lagrangian technique of ``Smooth particle hydrodynamics'' or SPH 
(Gingold \& Monaghan 1977; Lucy 1977). We focus on the thermodynamic
history of the gas that cools to make galaxies, on the rate at which
galaxies build up with time, on their abundance at the present
day and on their spatial and velocity distributions.  
Early results from one of the simulations that we analyse here were
presented in a previous paper (Pearce \etal 1999), where we focused on the
spatial distribution of galaxies.  Our work follows on from earlier
attempts to grow virtual galaxies in a cosmological volume using the SPH
technique, most notably by Carlberg, Couchman \& Thomas (1990), Katz \etal
(1992, 1996, 
1999), Evrard \etal (1994) and Frenk \etal (1996). These simulations
succeeded in resolving individual galaxies and allowed the first
investigations of their distribution and of the dynamics of galaxies in
clusters. However, the volumes modelled in these early simulations were
rather small, permitting only limited statistical studies. Our simulations
contain about one order of magnitude more particles and produce about 40
times more galaxies than the largest previous simulation of this kind by
Katz, Weinberg \& Hernquist (1996). SPH simulations with even higher
resolution have been used to study the formation of individual galaxies
(e.g. Navarro
\& Steinmetz 1997, Weil, Eke \& Efstathiou 1998).

Simulations of the kind studied here are complementary to semi-analytic
studies of galaxy formation (e.g. Kauffmann \etal 1997, 1999a,b; Diaferio
\etal 1999;  Guiderdoni \etal 1998; Benson \etal 1999a,b; Somerville \&
Primack 1999, Cole \etal 2000.) The main advantage of the hydrodynamic
simulations is that they follow the dynamics of diffuse cooling gas in full
generality whereas the semi-analytic treatment involves a variety of
approximations such as spherical symmetry and quasistatic evolution. A
comparison of results from the two methods has recently been performed by
Benson \etal (2000c). Both approaches require a phenomenological treatment
of critical processes such as star formation and feedback since these
involve scales far below the resolution limit of all current simulations.

The remainder of this paper is laid out as follows. In Section~2 we
describe our simulations and the physical approximations we have made. In
section~3 we discuss the evolution of the global properties of the gas, the
extraction of a dark matter halo catalogue, and the properties of the gas
within these halos.  In Section~4 we focus on the properties of galaxies,
their formation histories, luminosity function, age distribution, etc. In
Section~5 we consider the correlation function of the galaxies as a
function of redshift and galaxy mass, the relative pairwise velocity
distributions of galaxies and dark matter, and we determine how well the
galaxies trace the mass distribution within the largest dark matter halos.
We conclude in Section~6.  This work is part of the programme of the Virgo
consortium for cosmological simulations.

\section{Simulations}

\begin{figure*}
\psfig{file=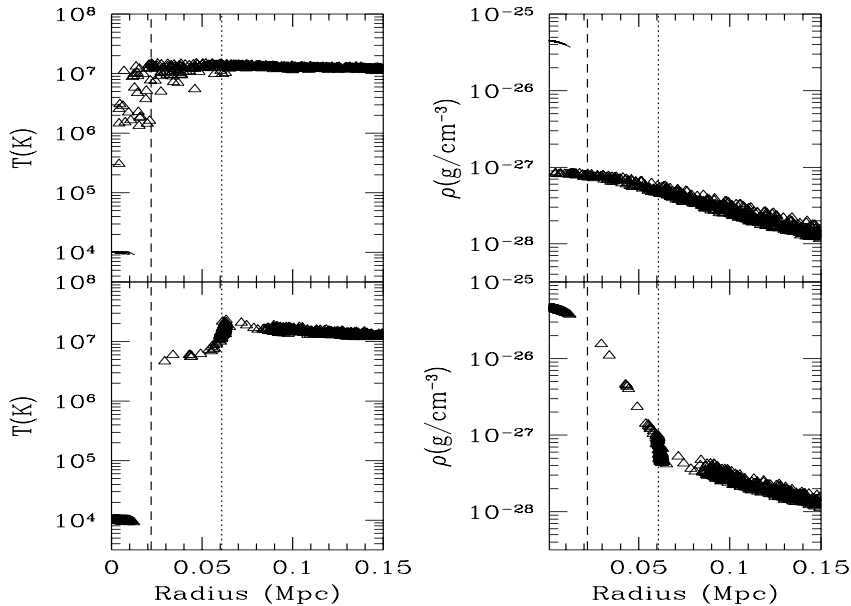,height=8.5cm,width=12cm}
\caption{An example of excessive cooling near a density interface. The two
panels on the left show radial temperature profiles and the two panels on
the right radial density profiles for a galaxy surrounded by a hot gaseous
corona. In both cases, the top panel shows the result of decoupling the
cold and hot gas phases as described in the text, while the bottom panels
shows the profiles with no decoupling of the cold and hot phases.  In the
latter case, the density associated with hot particles near the interface
between the two phases is overestimated, leading to enhanced cooling.  The
vertical dashed line shows the mean search length for particles which have
cooled into the galaxy and the vertical dotted line the mean search length
for the hot halo particles. In the case in which phase decoupling has been
implemented, the particles of the central galaxy are shown as dots rather
than triangles.  }
\label{supercool}
\end{figure*}

Our simulations were performed using a parallel, adaptive,
particle-particle, particle-mesh SPH code (Pearce \& Couchman
1997) which can be run in parallel on the Crays T3D and T3E, or
serially on a single processor workstation. This code is essentially
identical in operation to the publicly released version of {\sc
hydra}, described in detail by Couchman, Thomas \& Pearce (1995).

We have simulated galaxy formation in two cold dark matter (CDM) models
(e.g. Davis \etal 1985) with the same values of the cosmological parameters
as the SCDM and \LCDM models of Jenkins \etal (1998). For SCDM these are as
follows: mean mass density parameter, $\Omega_0=1$; cosmological constant,
$\Lambda/(3H_0^2)=0$; Hubble constant (in units of 100 km s$^{-1}$
Mpc$^{-1}$), $h=0.5$; and rms linear fluctuation amplitude in 8\hmpc
spheres, $\sigma_8=0.6$.  For \LCDM, the corresponding parameter values
are: $\Omega_0=0.3$; $\Lambda/(3H_0^2)=0.7$; $h=0.7$; and
$\sigma_8=0.9$. Each simulation followed 2097152 dark matter particles and
2097152 gas particles in a cube of side $100 \Mpc$ and required around
10000 timesteps (and $\sim10^5$ processor hours on a Cray-T3D) to evolve
from $z=50$ to $z=0$.  In both cases, we employed a $\beta$-spline
gravitational softening which remained fixed at the Plummer-equivalent
comoving value of $50\kpc$ until $z=1.5$ ($z=2.5$ for \LCDM). Thereafter,
the softening remained fixed at $50/(1+z)=10h^{-1}\kpc$ in physical
coordinates, and the SPH smoothing length was set to match this value.

We deliberately chose to set the gas mass per particle 
to be $\sim2\times10^9\Msun$ in
both cosmologies so as to have identical mass resolution. As 
we typically smooth over 32 SPH neighbours,
the smallest resolved objects have a gas mass of $\sim
6.4\times10^{10}\Msun$. 
The baryon fraction, $\Omega_b$, was set from the 
nucleosynthesis constraint, $\Omega_bh^2 = 0.015$ (Copi, Schramm \&
Turner 1995) and although this is somewhat lower than the more recent
limits given by Tytler \etal (2000), it was the current value at the
time these simulations were carried out.
We also assumed a constant gas metallicity of 0.3 times
the solar value in both cases.  We expect more gas to cool globally in
the \LCDM model because structure forms earlier in this model. With our
chosen parameters, our simulations were able to follow the cooling of
gas into galactic-sized dark matter halos. The resulting ``galaxies''
are typically made up of $50 - 1000$ particles and so have cold gas
masses in the range $10^{11}\Msun - 2 \times 10^{13}\Msun$. With a
spatial resolution of $10h^{-1} \kpc$, we cannot resolve the internal
structure of these galaxies and we must be cautious about the possibility of 
enhanced tidal disruption, drag, and merging within the largest
clusters. However, as we argue below, there is no evidence that this
is a major problem.

As in all numerical simulation work, approximations and compromises are
necessary if the calculation is to completed within a reasonable amount of
time. An important choice is the effective resolution of the
simulation. Increasing the spatial resolution requires increasing both the
number of particles (to prevent 2-body effects) and the number of timesteps
(to follow smaller structures). Our chosen value of $10h^{-1}\kpc$ for the
gravitational softening is larger than the scalelength of typical galaxies,
but is the best that we can do with our current computer resources. For a
given simulation volume, the number of particles determines the mass
resolution. In cosmological simulations that involve only collisionless
dark matter, the choice of particle number is often driven by computer
memory limitations.  Cosmological gas dynamics simulations are more
computationally demanding than collisionless simulations and memory
considerations do not, at present, limit their size. Since approximately
10\% of all the baryons in the universe have cooled by the present to form
stars and dense galactic gas clouds (Fukugita, Hogan \& Peebles 1998), and
around 50 particles are required to resolve a galaxy in a simulation, a
simple estimate of the number of gas particles required to obtain $N_{gal}$
galaxies is: $10\times N_{gal}\times 50$, or roughly {\it half a million}
particles for $1000$ galaxies. In practice, many galaxies contain more than
the minimum number of particles and so this is an underestimate. Since
galaxies of the characteristic luminosity, $L_*$ or greater, have a space
density of $\sim 0.01/(h^{-1}\Mpc)^3$, a simulation of a volume like those
studied by Jenkins \etal (1998), of side $\sim 250 h^{-1}\Mpc$, would
produce over $100,000 L_*$ galaxies and require around $10^8$ gas
particles. For this work, we have chosen to use just over $2\times 10^6$
million gas particles in a volume of side only 100 Mpc.

In addition to the choice of simulation parameters, other choices need to
be made to model physical processes that operate on sub-resolution
scales. Star formation and the associated feedback from energy released in
supernovae and stellar winds are the most obvious sub-resolution processes
relevant to simulations of galaxy formation. As the resolution 
is increased, progressively smaller, denser halos are resolved
whose gas has a short cooling time. In hierarchical clustering theories, a
simple model predicts that most of the baryonic material should have cooled
(and presumably turned into stars), in small, subgalactic objects at high
redshift (White \& Rees 1978, Cole 1991, White \& Frenk 1991). Since this
does not appear to be the case in the universe, some mechanism must have
prevented much of the gas from cooling. Reheating of the gas by the energy
released in the course of stellar evolution is a commonly invoked mechanism
to quench star formation in small galaxies. This, however, is still a very
poorly understood process. Various simplified models have been implemented
in simulations of galaxy formation (e.g. Katz 1992; Navarro \& White 1993;
Metzler \& Evrard 1994; Steinmetz \& Muller 1995; Evrard, Metzler \&
Navarro 1996; Katz \etal 1996; Mihos \& Hernquist
1994, 1996; Gerritsen \& Icke 1997; Navarro \& Steinmetz 1997).

In this paper, we have taken the simpler route of neglecting feedback
effects altogether. Instead, we appeal to the resolution threshold of the
simulation to prevent all the gas from cooling. This is analogous to
assuming a model in which gas cooling is suppressed within dark matter halos
of mass below $\sim 6.4\times10^{10}\Msun$), whilst leaving larger halos
unaffected. Although relatively simple, this model gives results that are
not too dissimilar from those produced by detailed semi-analytic models in
which feedback is treated with greater care.  In semi-analytic modelling, 
the star formation and feedback prescriptions are tuned by the requirement
that the model should reproduce observable properties of the local galaxy
population such as the luminosity function. The difference in the amounts
of gas that cool in our simulations and in the semi-analytic model of Cole
\etal (2000) are illustrated in Fig.~1 of Pearce \etal (1999).
Near the resolution threshold, approximately 50\% more gas cools in the
simulation than in the semi-analytic model and the difference falls rapidly
with increasing galaxy mass. A detailed comparison of results from these 
simulations and semi-analytic modelling may be found in Benson \etal
(2000c).

The second area in which sub-resolution physics is important is the
interaction of cool gas clumps with the surrounding hot gas. In reality,
the gas within such clumps would presumably become incorporated into
molecular clouds or turned into stars. These processes occur on scales that
are several orders of magnitude below the resolution limit of current
simulations and so must be modelled phenomenologically. Different authors
have chosen to do this in different ways. For example, Evrard \etal (1994)
and Frenk \etal (1996) opted to leave the cold gas alone. Although this is
the simplest possible approach, it has the disadvantage that the smoothing
inherent in the SPH method can artificially boost the cooling rate of hot
gas that comes into contact with the gas that has cooled to very high
density inside a galaxy. This process is illustrated in the lower panels
of Fig.~1 which show the temperature and density profile of a galaxy in the
simulation.  Model galaxies are typically surrounded by hot gaseous
coronae. At the interface with the cold galactic gas, the density of the
hot coronal gas is overestimated and its cooling rate is artificially
enhanced. This gives rise to the rapidly cooling material which is clearly
visible in the region between the effective search lengths of the cold and
hot material, indicated by the dashed and dotted lines in the figure
respectively. The artificial cooling at this contact interface arises
because SPH is not designed to deal with the steep density gradients that
develop in a multiphase medium.  As Thacker \etal (2000) have shown, this
limitation is present to some extent in all standard implementations of
SPH, although recently Ritchie \& Thomas (2000) have proposed an
alternative formulation of SPH utilising the gas pressure rather than
the gas density which overcomes this limitation.

\begin{figure}
\psfig{file=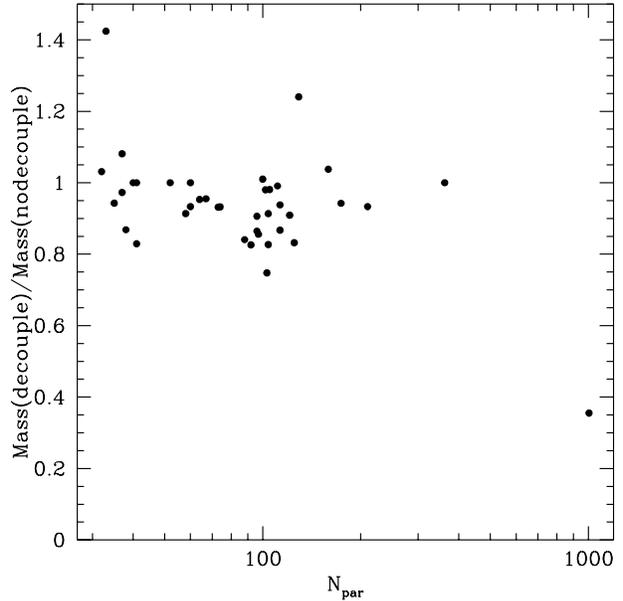,height=8.5cm,width=8.5cm}
\caption{The effect of decoupling the hot halo gas from the cold
galactic material. The graph shows the ratio of the masses of matched
clumps found in two test simulations with similar parameters and resolution
to those assumed in our main simulations. In one of the test simulations,
the hot gas was decoupled from the cold gas, as described in the text; in
the other it was not. The quantity plotted along the $y$-axis is the ratio
of the clump mass in the first to the clump mass in the second of these
test simulations.  Only the largest object (which contains over 1000
particles in the simulation without decoupling) is significantly affected.
\label{cool}}
\end{figure}

An alternative approach to leaving the galactic cold dense gas alone
is to replace it with collisionless ``star'' particles, thus
decoupling it entirely from the rest of the baryonic material
(Navarro \& White 1993, Steinmetz \& Muller 1995, Katz \etal
1996). While this might appear as a more realistic solution than
leaving cold material as gas, in practice, the relatively large
softening lengths normally used in SPH simulations imply
unrealistically low binding energies for the star particles. As a
result, stellar objects are prone to disruption by collisions and
tidal encounters, particularly in strongly clustered regions.

In this study, we have chosen an approach which is intermediate between the
two extremes just discussed. In our phenomenological model, hot gas,
defined to be hotter than $10^5\K$, is prevented from interacting with
(galactic) material at a temperature below $12000\K$ (but not
vice-versa). All other SPH forces
remain unchanged. The threshold temperature of $10^5\K$ is less than the
virial temperature of the smallest resolved halos in our simulations. In
our implementation of SPH, this assumption preserves force symmetry and
momentum conservation, and effectively prevents overcooling of hot gas
induced purely by the presence of a nearby clump of cold, dense gas. This
is clearly seen in the top panels of Fig.~1. In contrast to the standard
model in which the hot and cold phases remain strongly coupled, there is
now no interface discontinuity. As a result, the density profile in the
vicinity of a cold clump remains smooth and a ``cooling flow'' develops
around it. Recently this effect has been confirmed by Croft \etal
(2000) who suggest several alternative routes to circumvent it.

\begin{figure}
\psfig{file=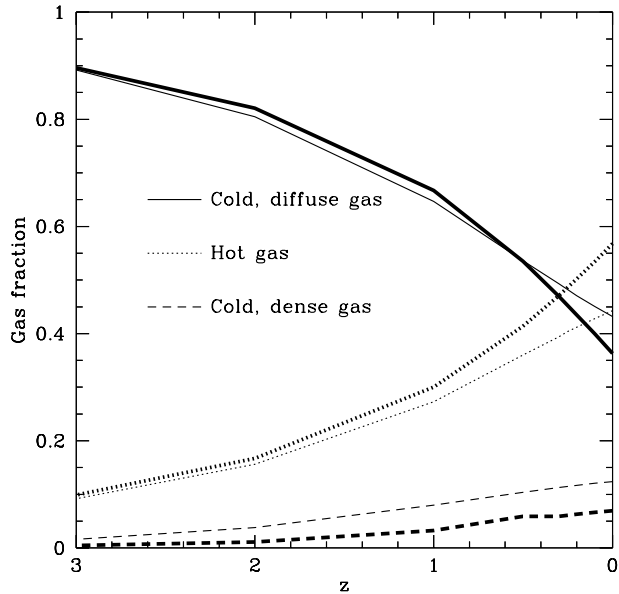,height=8.5cm,width=8.5cm}
\caption{The relative fractions of gas in each of three phases: (i)
cold, galactic gas (temperature less than $12000\K$, overdensity greater
than 10); (ii) hot halo gas (temperature greater than $10^5\K$); (iii) cold,
diffuse gas in ``voids'' (overdensity less than 10).  
The bold lines correspond to the SCDM simulation and the thin lines to the 
\LCDM simulation. 
\label{gastemp}}
\end{figure}

In our scheme, the cold gas ``feels'' all the usual SPH forces, including
viscous drag when it moves in a hot intergalactic medium, and interacts
with other cold gas clouds in the usual way. The outer accretion shock of a
galaxy cluster is still properly modelled as the infalling material can
``see'' the hot gas in the cluster. Our approximation has only a small
effect on the masses of small galactic objects as may be seen in Fig.~2,
where we compare the masses of galaxies formed in two small test
simulations. These had similar parameters and mass resolution to those
employed in our main simulations, but in one case the hot gas was decoupled
from the cold gas and in the other case it was not.  For most
galactic-sized objects, this change produced a slight reduction in mass,
but for the largest object, which has over 1000 particles without
decoupling, the mass is considerably reduced to 370 particles. This effect,
as well as the results of varying several simulation parameters are
explored further by Kay \etal (2000).

\section{The properties of the gas}

In this section, we examine both the evolution of the gas in the simulation
as a whole and the properties of gas within virialised dark matter halos.

\subsection{Global properties of the gas} 

Following Kay \etal (2000), we define three different gas phases: (i)
cold galactic gas (temperature below $12000\K$ and overdensity greater
than 10); (ii) hot halo gas (temperature above $10^5\K$); and (iii)
cold, uncollapsed gas (overdensity less than 10; this is essentially
diffuse gas in ``voids.'') We ignore the effects of photoionization
because at this mass resolution a photoionising field has a negligible
effect upon galactic or halo properties.
In practice, little gas is both cold and at overdensities between 5
and 1000, so that effectively the uncollapsed and galaxy phases are
completely disjoint. The evolution of the fraction of gas in each of
these phases is illustrated in Fig.~3 for the two cosmological models
that we have calculated, SCDM (bold lines) and \LCDM (thin
lines). These gas fractions are very similar to those derived using a
semi-analytic model of cooling gas implemented in these simulations by
Benson \etal (2000c).

\begin{figure}
\psfig{file=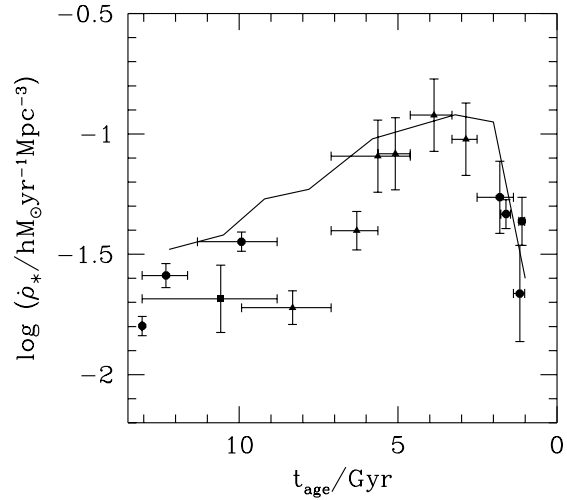,height=8.5cm,width=8.5cm}
\caption{The rate at which gas cools, per unit volume, in the SCDM
simulation as a function of time since the Big Bang. 
The data points are inferred star formation rates in the real
universe, from Steidel \etal (1999). The match between the simulation
results and the data indicates that, in spite of resolution effects and
other limitations, the simulations provide a reasonably realistic model. 
\label{madau}}
\end{figure}

The properties of cold, uncollapsed material in voids are only followed in an
approximate manner in our simulations. The SPH method generally has
poor spatial resolution in regions in which the gas is diffuse because
it must smooth over a sparse particle distribution.  This limitation
is exacerbated in our specific implementation which imposes a 
maximum search length, resulting in a minimum resolvable density of a
few times the mean density in the computational volume.  Test
simulations without such a minimum resolvable density demonstrate that
these inaccuracies do not affect the properties of the galaxies because
the gas forces are small in void regions and have no effect on the
dense regions that ultimately form galaxies (Kay \etal 2000). However,
outwardly propagating shocks generated by the formation of large
clusters are poorly followed in underdense regions and so we can say
little about the detailed properties of voids (beyond their mere
presence). This is a regime in which Eulerian techniques are more
accurate than SPH (\eg Cen \& Ostriker 1996).

As the evolution proceeds the cold, diffuse phase is depleted as gas
falls into dark halos where it is shocked into the hot phase,
thereafter cooling into galaxies.  The time derivative of the dashed
curve that tracks the galaxy phase in Fig.~3 effectively defines the
global star formation rate in the computational volume. This is
plotted in Fig.~4. The star formation rate rises rapidly at early times, has a
broad maximum around $z\sim 1-2$ and then gently declines to the
present. This behaviour is broadly similar to that seen in real data
(Steidel \etal 1999), indicating that the limited resolution of the
simulations and other approximations do not result in too unrealistic
a model. At late times, Fig.~3 shows that there is a substantial
transfer of gas from the uncollapsed phase into the hot phase, with
around $50\%$ of the gas ending up at temperatures above $10^5\K$ in
both cosmological models.  At $z\simeq 1$, when only around $30\%$ of
the gas is above $10^5\K$, the gas fractions in the different phases
in our simulations agree well with those in the simulations of Evrard
\etal (1994) (which were stopped at this point.)

\begin{figure}
\psfig{file=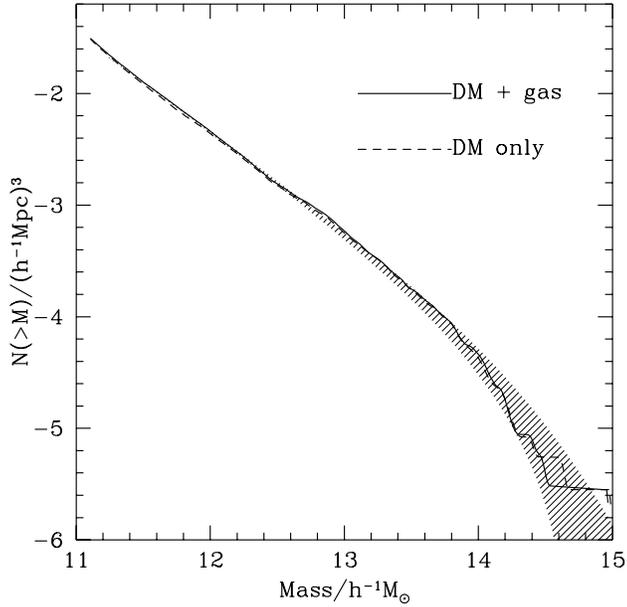,height=8.5cm,width=8.5cm}
\caption{A comparison of the cumulative mass function of dark
halos in the \LCDM gas + dark matter simulations and an otherwise
identical pure dark matter simulation. The halo masses of the gas +
dark matter simulation have been scaled by
$\Omega_0/(\Omega_0-\Omega_b)$.  For this plot the halos were
identified using a friends-of-friends algorithm with a linking length
of 0.2. The shaded region shows the expected number of halos, allowing
for $1\sigma$ Poisson errors because of the finite simulation volume, as
predicted by the formula given in Jenkins \etal (2000).
\label{ps}}
\end{figure}

As expected, more gas cools into the galaxy phase in the \LCDM than in
the SCDM model. This is due to our adoption of a single gas particle
mass in the two simulations which results in the same effective
resolution in both cases. More gas cools at early times in the \LCDM
simulation because structure forms earlier in this case. If all the
gas that cools into galaxies were assumed to turn into stars, then the
SCDM model would have roughly the observed mean mass density in
stars. In the \LCDM model, on the other hand, about twice as much gas
cools and, if all this gas were assumed to turn into stars, an
uncomfortably large value of the mass-to-light ratio would then be
required to match the observed mean stellar mass density. The amount
of gas that cools in our simulations is determined by the mass
resolution and a small decrease in the resolution of the
\LCDM model would reduce the final amount of cold gas to the same
level as in the SCDM model.

\subsection{Properties of gas in halos}

To construct a catalogue of dark matter halos from the simulations, we
first identified cluster centres using the friends-of-friends grouping
algorithm (Davis \etal 1985) with a small value of the linking length,
$b=0.05$. We then grew a sphere around the centre of mass of each of these
halos until the mean overdensity within it reached a value of 178 in the
case of SCDM or 324 in the case of \LCDM. These are the overdensities
corresponding to the virial radius according to the spherical top-hat model
for clusters (Eke, Cole \& Frenk 1996.)  The resulting set of halos was
cleaned by removing the smaller of any two overlapping halos.  This
procedure produced catalogues of 1320 halos in SCDM and 1795 halos
in \LCDM, with more than 28 particles per halo in each case. 

To check whether the presence of baryons affects the halo catalogue, we ran
a pure N-body simulation of the \LCDM model and extracted a halo catalogue
from both it and the baryonic simulation with gas cooling. The mass
function of halos in the test simulation is almost indistinguishable from
that in the original simulation, as illustrated in Fig.~5. This figure also
compares the mass functions in the simulations with the mass function given
in Jenkins \etal (2000). This gives a better fit than the predictions of
the Press-Schechter model (Press \& Schechter 1974).  Although for the
values of $\Omega_b$ that we have assumed here, the presence of gas has
little effect on the halo mass function, it does affect the central
structure of the halo into which it cools (Pearce \etal 2000).

\begin{figure*}
\psfig{file=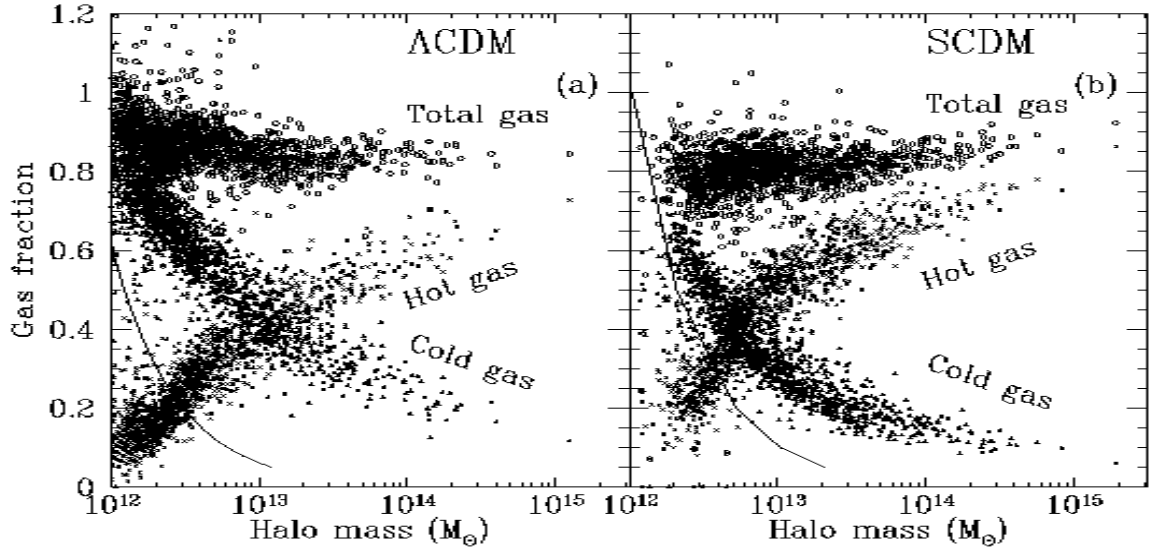,height=8.5cm,width=17.5cm}
\caption{The ratio of gas to dark matter mass ($\times
\Omega_{cdm}/\Omega_b$, where $\Omega_{cdm}$ denotes the dark matter
contribution to $\Omega_0$) within the virial radius of $\sim 1500$
dark matter halos found in the simulations. The circles give the total gas
fraction, the triangles the fraction of gas below $12000\K$ and the crosses
the fraction of gas above $12000\K$.  The solid line indicates the
effective resolution limit of 32 gas particles within a halo. Panel (a)
corresponds to the \LCDM simulation and panel (b) to the SCDM simulation. 
\label{frac}}
\end{figure*}

\begin{figure*}
\psfig{file=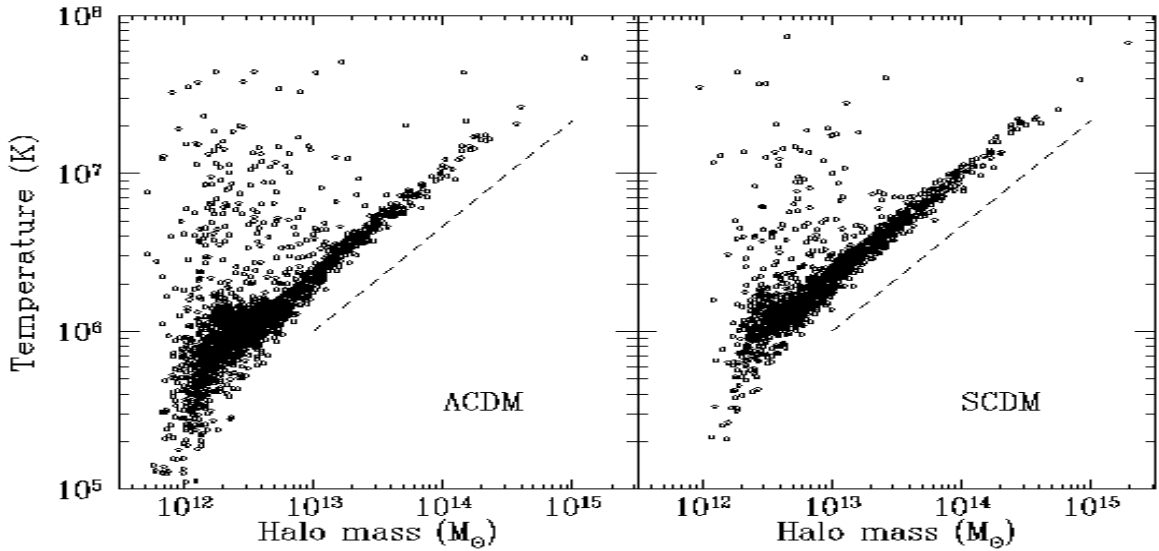,height=8.5cm,width=17.5cm}
\caption{The relation between the mean, mass-weighted temperature of
hot ($>10^5\K$) gas and the mass of its host dark matter halo.  The
dashed line shows the theoretical expectation, $M \propto T^{3 \over
2}$, of a simple equilibrium model for the halos. The tight
correlation between $T$ and $M$ indicates that the gas is close to
equilibrium in all but the smallest halos in the simulations (where
the temperatures can be overestimated due to the proximity of a larger
halo, see text).  
\label{hgvstemp}}
\end{figure*}

In Fig.~6 we plot the mass fraction of gas within the virial radius of the
halos normalized to the mass fraction in the simulation as a whole,
together with the proportions in the hot and cold phases. The total gas
mass fraction varies little with halo mass over the range $10^{12}\Mo\lsim
M\lsim 10^{15}\Mo$, although there is some indication of an upturn at low
masses in the \LCDM case and at high masses in the SCDM case. The mean
value of the gas fraction is $\sim 0.85$ for \LCDM and $\sim 0.8$ for SCDM
with a scatter of $\sim 0.05$, indicating that the gas is slightly more
extended than the dark matter in both models. A few small mass halos have a
gas mass fraction greater than unity. These halos, however, are near the
resolution limit of the simulations.  The crosses show the mass fraction of
hot ($T> 12000\K$) gas in the halos. This increases rapidly with halo mass,
from $\sim 10-20\%$ for the smallest halos in the simulations to $\sim
60-80\%$ for the largest. The fraction of cold ($T<12000\K$) gas shows the
opposite trend and varies from $\sim 70-80\%$ at the small mass end to
$\sim 10-20\%$ at the high mass end.

\begin{figure*}  
\psfig{file=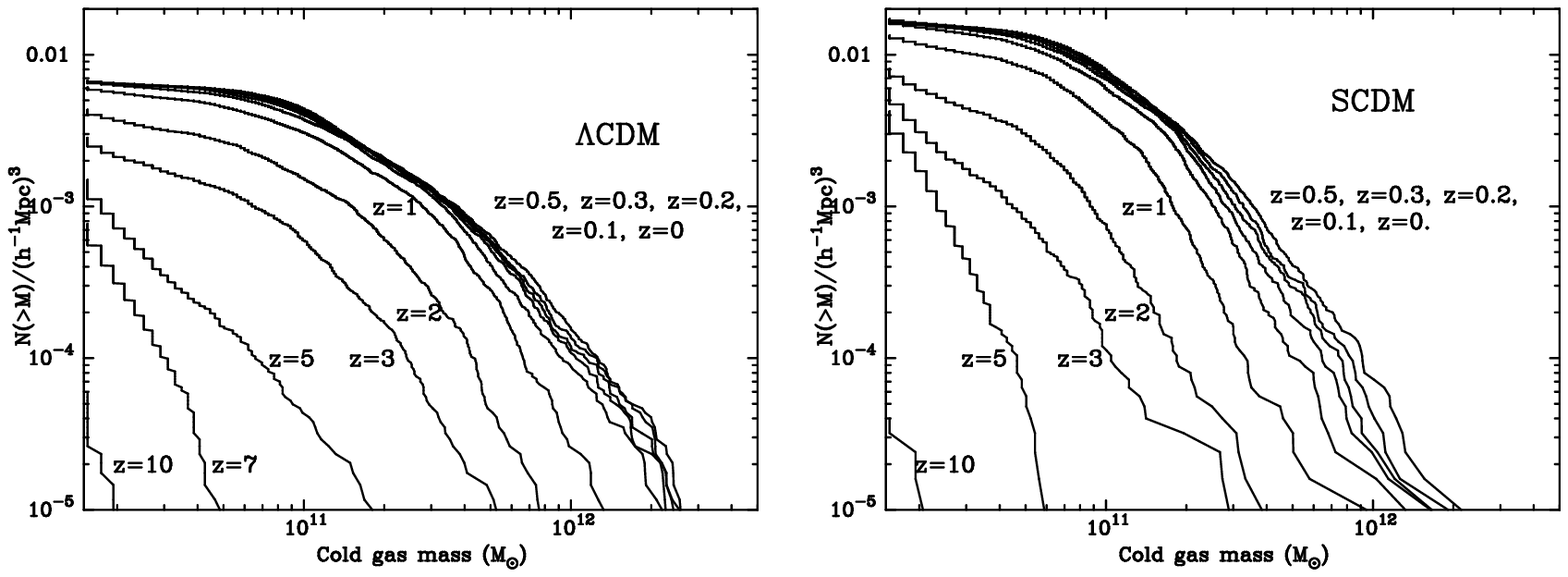,height=8.5cm,width=15.0cm}
\caption{Evolution of the abundance of ``galaxies" in the simulations as a 
function of their cold gas mass. The mass functions are plotted at
the redshifts given in the legend. The left panel shows results for \LCDM
and the right panel for SCDM.
\label{mult}}
\end{figure*}

The fraction of baryonic material that cools within dark matter halos can
also be calculated using semi-analytical techniques (White \& Frenk
1991). As we discussed in the preceeding section, in semi-analytic models
the cooling of gas is regulated by feedback processes due to energy
liberated by stars. In these simulations we ignore feedback altogether and
cooling in small mass objects is then limited by resolution effects. This
leads to some differences in the gas fractions derived with these two
techniques. For example, in the semi-analytic model of Cole \etal (2000),
the gas fraction of cold gas has a maximum of $\sim 40$\% at a halo mass of
$10^{12}\Msun$. This is close to the resolution limit of our
simulations. At small masses, the cold fraction is suppressed by feedback
and at larger masses, it is suppressed by the long cooling time of the
diffuse gas in halos. The first of these effects is not present in our
simulations (which do not resolve the regime where the feedback is
important in the semi-analytic models), but the second one is. On these
large mass scales, the simulations agree well with the semi-analytic
results and even over the entire range of mass, the difference is only of
the order of 50\% (Benson \etal 2000c). Overall, somewhat more mass cools
in the simulations than in the semi-analytic model, suggesting that
feedback effects are not negligible even above the resolution limit of our
simulations.

The hot gas that falls into a halo is quickly thermalized by shocks and
heated to the virial temperature. This is illustrated in Fig.~7 which shows
the relation between the mean, mass-weighted gas temperature (for gas with
$T>10^5\K$) and the virial mass of the host halo. For masses above a few
times $10^{12}\Msun$, there is a very tight correlation between gas
temperature and halo mass, indicating that the gas is relaxed and close to
equilibrium. At the small mass end, there are some halos with anomalously
large temperatures. These tend to be in the vicinity of larger halos whose
own hot halos contaminate the smaller ones, boosting their temperature.
The dashed line is the relation $M \propto T^{3
\over 2}$ predicted by a simple, spherically symmetric, equilibrium
model for the halo and its gas. The simulations follow this relation
well over most of the mass range plotted in the figure. The X-ray
properties of the largest halos in our simulations have been studied by Pearce
\etal (2000).

\section{The properties of galaxies}

We now consider the properties of the population of galaxies in the
simulations, focussing on their abundance, ages, merger histories, and 
individual star formation rates. 

\subsection{Demographics of the galaxy population} 

We identify ``galaxies" in the simulations with dense clumps of cold
gas. Locating them is straightforward except in the small
number of cases when they are undergoing a merger or are being tidally
disrupted within a large cluster. To find galaxies, we use the same
friends-of-friends algorithm that we employed for finding dark halos,
but with a linking length of only $b=0.02$ for SCDM and $b=0.0134$ for
$\Lambda$CDM, values that are 10\% of those required to obtain virialised
halos in these cosmologies.
When calculating the mass of each galaxy, we consider only particles 
with temperature below $12000\K$, a condition that
rejects only very few particles. In practice, as the cold clumps of
gas that make up the galaxies have contracted by a large factor the 
galaxy catalogues and the masses
and positions of each are insensitive to these choices.  We find a
total of about 2000 resolved galaxies in each of the two simulations at the
present day.

The evolution in the abundance of galaxies is shown in Fig~8. Here, we
characterize each galaxy by its cold gas mass and plot the cumulative
abundance at various epochs in our two simulations. In the \LCDM model, the
first substantial galaxies begin to form at a redshift $z\simeq 5$,
although some fairly massive objects are present even before this. The
massive galaxy population builds up rapidly between $z\simeq 5$ and
$z\simeq 3$ when it increases approximately ten-fold. A similar increase
occurs between $z\simeq 3$ and $z\simeq 1$. As time progresses, the mass of
the largest galaxies also increases. After $z=0.5$, the growth in the
population slows down somewhat because the rate at which new galaxies are
forming is more or less balanced by the rate at which existing galaxies are
destroyed by mergers and tidal disruption. At $z=0$, the abundance of
galaxies with gas mass $M>10^{11} \Msun$ approximately matches the observed
abundance of $L_*$ or brighter galaxies of $\sim 0.003$ per
$h^{-3}\Mpc^{3}$. The evolution in the SCDM model is qualitatively similar,
but shifted towards lower redshifts.

The creation sites of galaxies at recent times are 
illustrated in Fig.~9 which shows the locations where new
gas cools between $z=0.3$ and the present in the SCDM model. (Broadly similar
behaviour is seen in the \LCDM model.) The radius of the circle around
each of these locations is scaled by the mass of new cold
material. Some gas cools onto preexisting objects (note the small
circles at the centre of some clusters), but 383 new resolved
galaxies, about 15\% of the total, formed during this
period. The new galaxies are
small and tend to form along filaments or in low density regions, away
from the larger clusters. 

\begin{figure}
\psfig{file=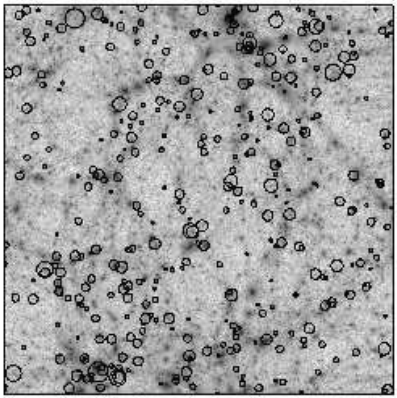,height=8.5cm}
\caption{The circles indicate the 
locations where new gas cools after $z=0.3$ in the
SCDM model, scaled in radius by the mass of new
cold material.  The grey shading indicates 
the underlying dark matter density at $z=0$.
\label{newgal}}
\end{figure}

Galaxies are destroyed preferentially in large
clusters, where tidal forces are largest.
For example, about $20\%$ of the cold material within the virial
radius of the largest halo at $z=0.3$ is reheated before $z=0$, and is
associated with the destruction of about half the small (around
1/3$L_*$) galaxies in this halo. As discussed earlier,
tidal disruption of galaxies in clusters is likely to be enhanced by
the limited spatial resolution of the simulations which results in 
artificially low binding energies for small objects. Larger simulations 
are required to investigate this important problem in detail. 

The major consequence of ``decoupling'' the hot gas from the cold
phase, as discussed in the preceeding section, is that no supermassive
objects form. The biggest galaxy in the \LCDM model contains $5\times
10^{12} \Msun$ of cold gas.  The turnover in the abundance of galaxies
at small masses in the two simulations is a numerical artifact. At
least 32 particles (corresponding to a mass of $6.4\times10^{10}
\Msun$) of gas must be present for cooling to be efficient; objects
below this are in the process of forming or are being ablated as they
move in a hot halo. The effective mass threshold in
both simulations is actually slightly higher than the nominal value
and corresponds to a mass threshold around $10^{11}\Msun$, or 50
gas particles.

The number of galaxies within the virial radius of each halo at the present
day is plotted against the halo mass in Fig.~10. The largest cluster
contains nearly 30 large galaxies and has a virial radius of over
$1.5h^{-1}\Mpc$. Its galaxy content is consistent with those of $10^{15}
h^{-1} \Msun$ clusters which typically contain 30 - 100 $L_*$ galaxies.
This is, in fact, the only large Abell-type cluster that formed in our
simulations.  All dark halos in the simulations that contain more than a
few hundred particles have at least one galaxy within them. There are no
{\it empty halos} of mass greater than $10^{13}\Msun$ in either cosmology.

\begin{figure*}
\psfig{file=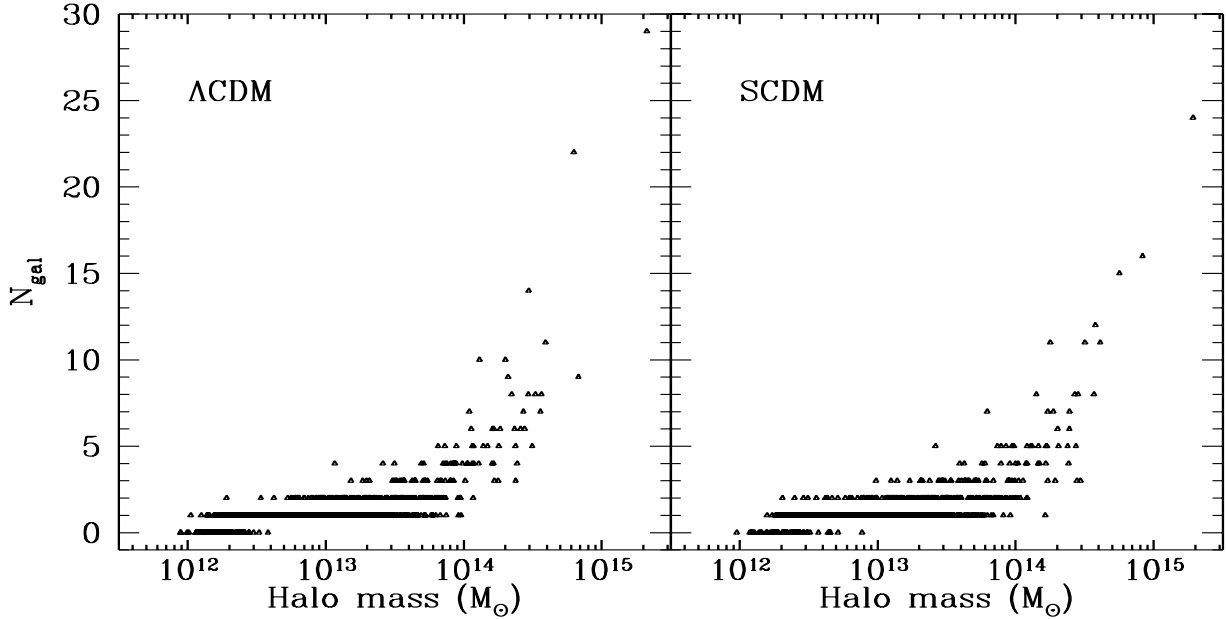,height=8.5cm,width=17.5cm}
\caption{The number of galaxies with mass above
$6.4\times10^{10}\Msun$ found within the virial radius of each dark
matter halo. In the \LCDM model, the largest halo contains 29 galaxies
(24 in SCDM) and there are 6 clusters of 10 or more galaxies (7 in
SCDM). The largest clusters have suffered
significant depletion of their galaxy population due to ablation of
galaxies moving through the cluster halo.
\label{ngal}}
\end{figure*}

\subsection{Star formation and the age of galaxies}

Although, for simplicity, we have not included any prescription for star
formation in our simulations, we can still derive some general conclusions
regarding the properties of the stars expected to form. Stars form from gas
that has cooled into galactic dark halos. The approximate time at which an
individual gas particle is available for star formation is easy to estimate
in the simulation. We take this to be halfway between the output time at
which the particle first has an overdensity greater that 10 and a
temperature smaller than $12000\K$, and the previous output time. These
criteria have previously been shown to pick out solely baryonic material
that has cooled into galactic objects (e.g. Kay \etal 2000). We assume that
as soon as a particle becomes available for star formation, a mass of stars
equal to the mass of the particle forms. 

By calculating a star formation time for each particle that ends up in
galaxy, we can derive a mean, mass-weighted stellar age or formation time,
$t_{\rm form}$, for each galaxy as
\begin{equation}
t_{\rm form} = \sum_{i=1}^{N} \frac{t_i}{N}, 
\end{equation}
where the sum is over all particles found within the final galaxy.  This
definition of formation time is based exclusively on the age of the stellar
population of a galaxy and takes no account of when the galaxy was
assembled.  An alternative definition of formation epoch is the time at
which half the final galaxy mass was assembled into a single progenitor. As
illustrated in Fig.~11, these two times can be very different. For the
largest galaxy in the SCDM simulation volume (which has a gas mass of $2.4
\times 10^{12}\Mo$ and for which we have enough resolution to follow many
progenitors), the average formation time of its stars corresponds to
redshift $z\simeq 2.5$, but half of the stellar material is only assembled
into a single distinct object at $z\simeq0.7$.  The formation history of
this galaxy bears little resemblance to the classic spherical top-hot
collapse model. Rather, the galaxy is assembled through repeated mergers of
sub-units, many of which have a mass close to the resolution threshold of
the simulation and which are originally spread over quite a large comoving
volume. These fragments tend to line up with filaments of the dark matter
distribution, giving rise to an anisotropic fragment distribution which is
particularly visible at $z=3$ in Fig.~11. The mass distribution of the
progenitors of this massive galaxy at various epochs is shown in Fig.~12.

\begin{figure*}
\psfig{file=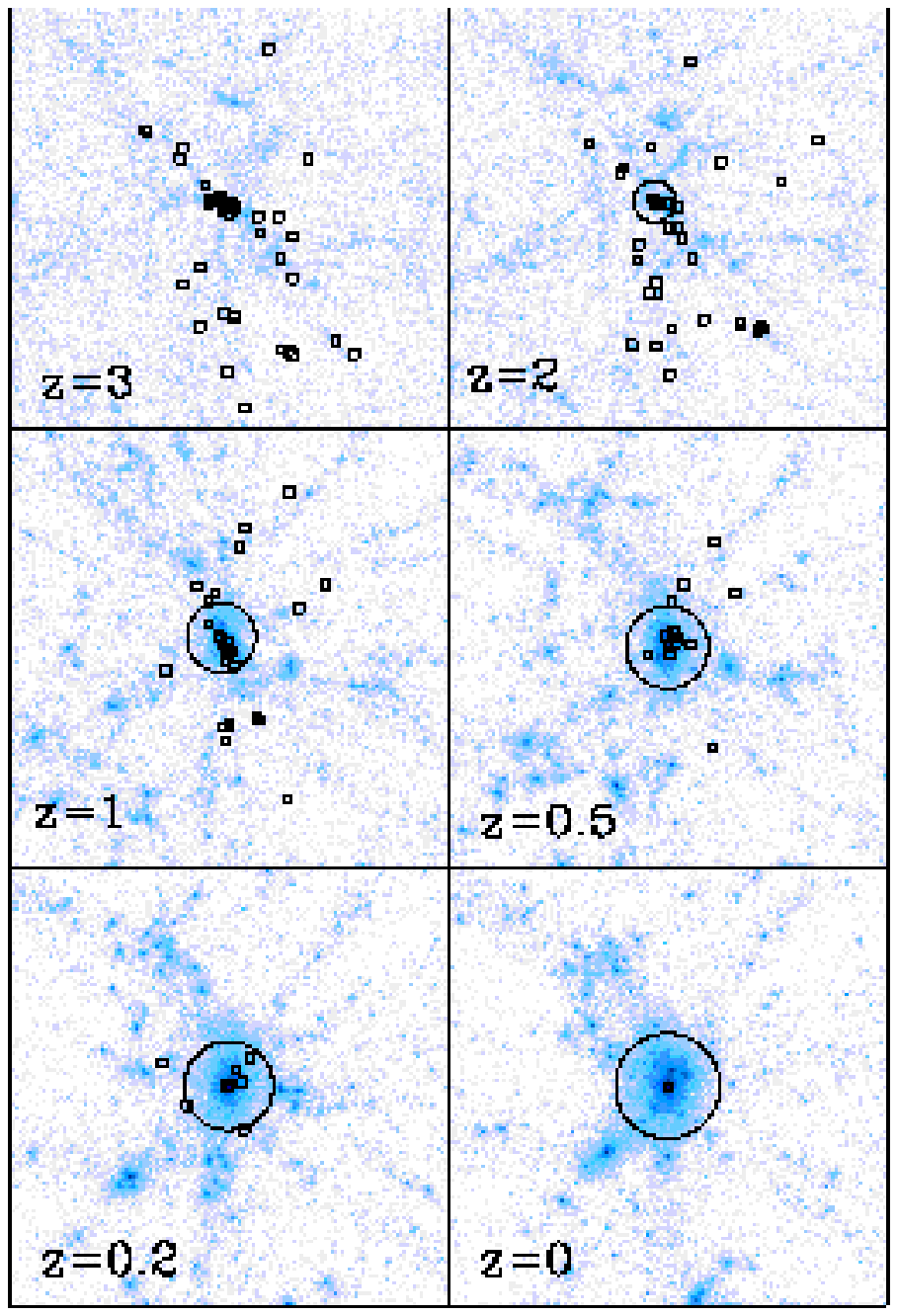,height=22.5cm}
\caption{The positions of the progenitor fragments that make up
the largest galaxy at $z=0$ in the SCDM cosmology.  Objects containing more
than 10 particles are shown as open squares. At each redshift, the circle
indicates the virial radius of the largest progenitor of the dark matter
halo hosting the largest galaxy, and the grey shading shows the underlying
dark matter density. Each frame illustrates a comoving cubic region of side
$12h^{-1}\Mpc$.  Half the mass has cooled by $z\simeq 2.5$, but it is not
assembled into a single object until around $z\simeq 0.7$.  
\label{galform}}
\end{figure*}

\begin{figure}
\psfig{file=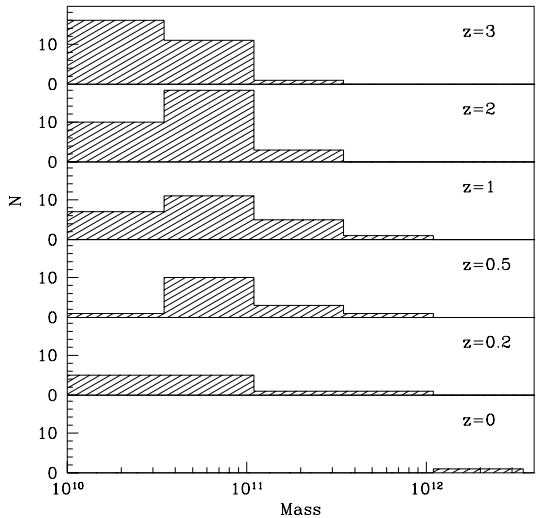,height=7.4cm,width=7.4cm}
\caption{The mass distribution of progenitors of the most massive
galaxy in the simulation. Each histogram refers to a different redshift as
shown in the legend.
\label{progenitor}}
\end{figure}

Fig.~13 shows the distribution of the mean stellar ages (as a fraction of
the lookback time) for galaxies in the SCDM simulation.  Contrary to the
naive expectation for a hierarchical scenario, there is a weak trend for
the largest galaxies to be the oldest. This may be an artificial
consequence of the resolution limit of our model which is at a fixed
mass. Models with more realistic feedback and no lower mass limit produce a
weak trend in the opposite sense for galaxies in the field and no trend at
all for galaxies in clusters (Kauffmann 1996).

\begin{figure}
\psfig{file=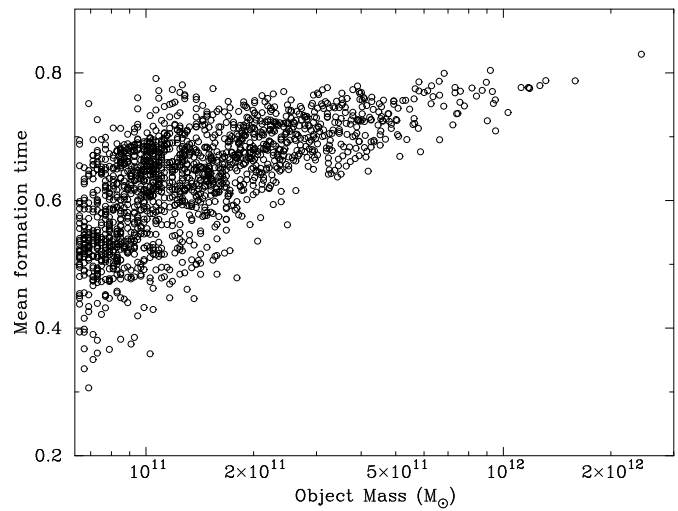,height=8.0cm}
\caption{The age distribution of mean stellar ages of galaxies in
the SCDM simulation. The age is shown as a
fraction of the lookback time (i.e. 0 corresponds to the present and 1
to the Big Bang). The largest galaxies are the oldest because their
stars formed preferentially in small fragments at  
at high redshift (see fig.~11).
\label{age}}
\end{figure}

\begin{figure}
\psfig{file=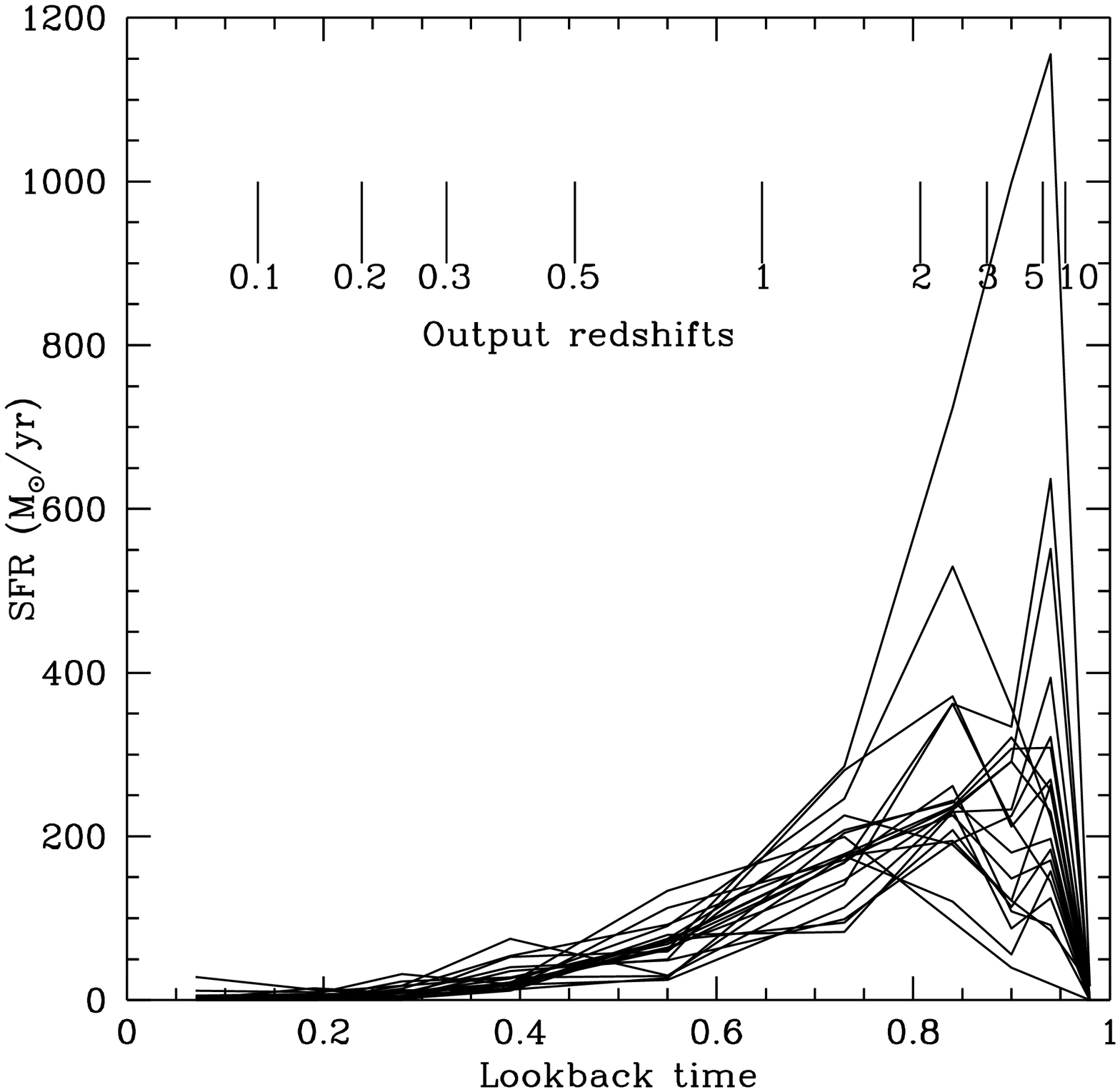,height=8.5cm,width=8.5cm}
\caption{The star formation rate for the 20 largest galaxies by
mass selected at $z=0$ in the SCDM simulation as a function of cosmic
time.
\label{bigsfr}}
\end{figure}

An estimate of the star formation rate (SFR) in our simulated galaxies can
be derived from the gas mass that cools between successive output
times. This crude measure (we only have 10 output times) is shown for the
20 largest galaxies selected at the end of the SCDM simulation in
Fig.~14. Clearly, the inferred aggregate SFR within each of these objects
is very high but, as Fig.~11 shows, these galaxies are broken up into many
small precursor objects at early times each of which has a modest SFR. The
aggregate SFR for the most massive objects today is sharply peaked at
$z\simeq 3$.  This contrasts with the star formation history of more
typical galaxies, illustrated in Fig.~15. This shows SFRs for every
$50^{\rm th}$ galaxy ordered by mass. These curves are more extended in
time and the mean SFR for the average galaxy is much lower than the
aggregate star formation rate of the largest objects. The SFRs generally
decline towards low redshift as galaxies consume the gas that is able to
cool by the present day. The SFR integrated over all the galaxies in the
simulation is shown in Fig.~4.

\begin{figure}
\psfig{file=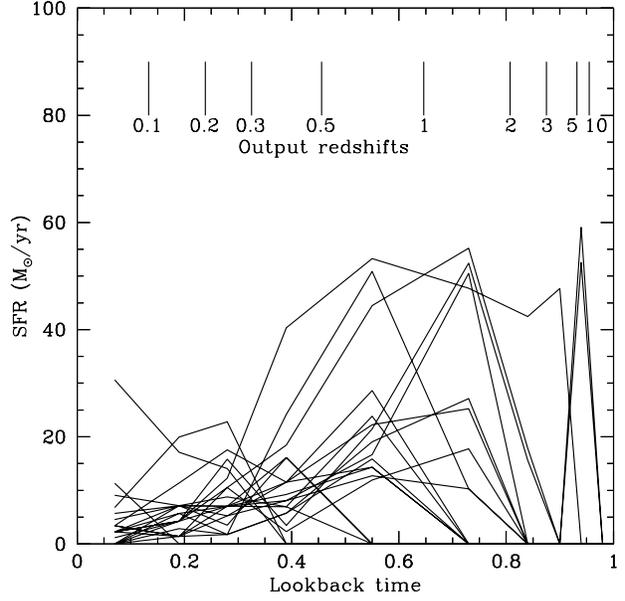,height=8.5cm,width=8.5cm}
\caption{The star formation rate for every 50th galaxy ordered
by mass selected at $z=0$ in the SCDM simulation as a function of
cosmic time. 
\label{allsfr}}
\end{figure}

\begin{figure*}
\psfig{file=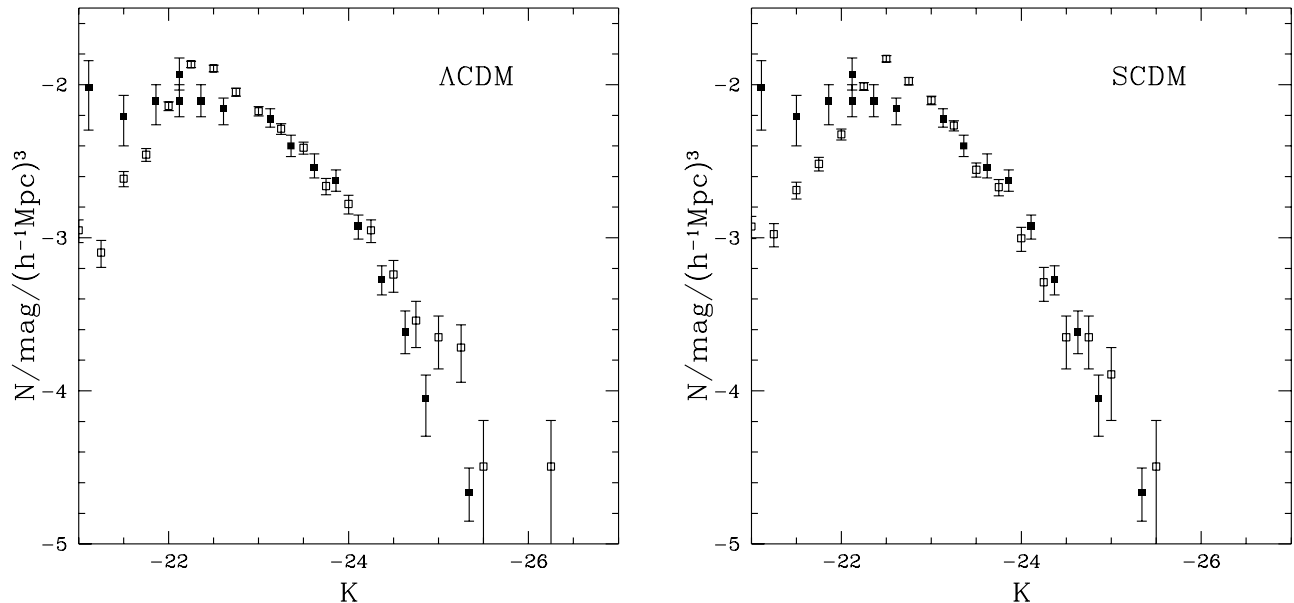,height=8.5cm,width=17.5cm}
\caption{The $K-$band luminosity function. The left panel shows 
the \LCDM model and the right panel the SCDM model. The open
squares show our model results, while the filled squares are
observational data from Gardner \etal (1997). Poisson error bars are
shown in both cases.
\label{kband}}
\end{figure*}

By convolving the SFR in each galaxy with the Bruzual \& Charlot (1993)
stellar population synthesis model, we have calculated the $K-$band
luminosity of each galaxy.  We have assumed a Salpeter Initial Mass
Function with a low mass cutoff of $0.1\Msun$ and solar metallicity. (This
is higher than our assumed metallicity of 0.3 solar for the ambient gas,
since the galactic material will have been enriched by reprocessed metals.)
As shown in Fig.~16, the model luminosity function in both models has a
similar shape to the $K-$band luminosity function measured by Gardner \etal
(1997). Furthermore, the model luminosity function can be made to agree
quite well with the data by assuming that only a fraction $1/\Upsilon$ of
the stellar mass is luminous (and the rest is in brown dwarfs), with
$\Upsilon =2.8$ for \LCDM\ and $\Upsilon=0.83$ for SCDM. The former is
somewhat larger and the latter somewhat smaller than the values inferred
from observations (Cole \etal 2000 and references therein) because, as
discussed earlier, we have used the same mass per gas particle in the two
simulations and, as a result, more gas cools globally in the \LCDM model
than in the SCDM case. Due to limited resolution, only the bright end of
the luminosity function ($L \gsim L_*$) is accessible in our simulations.

\begin{figure*}
\psfig{file=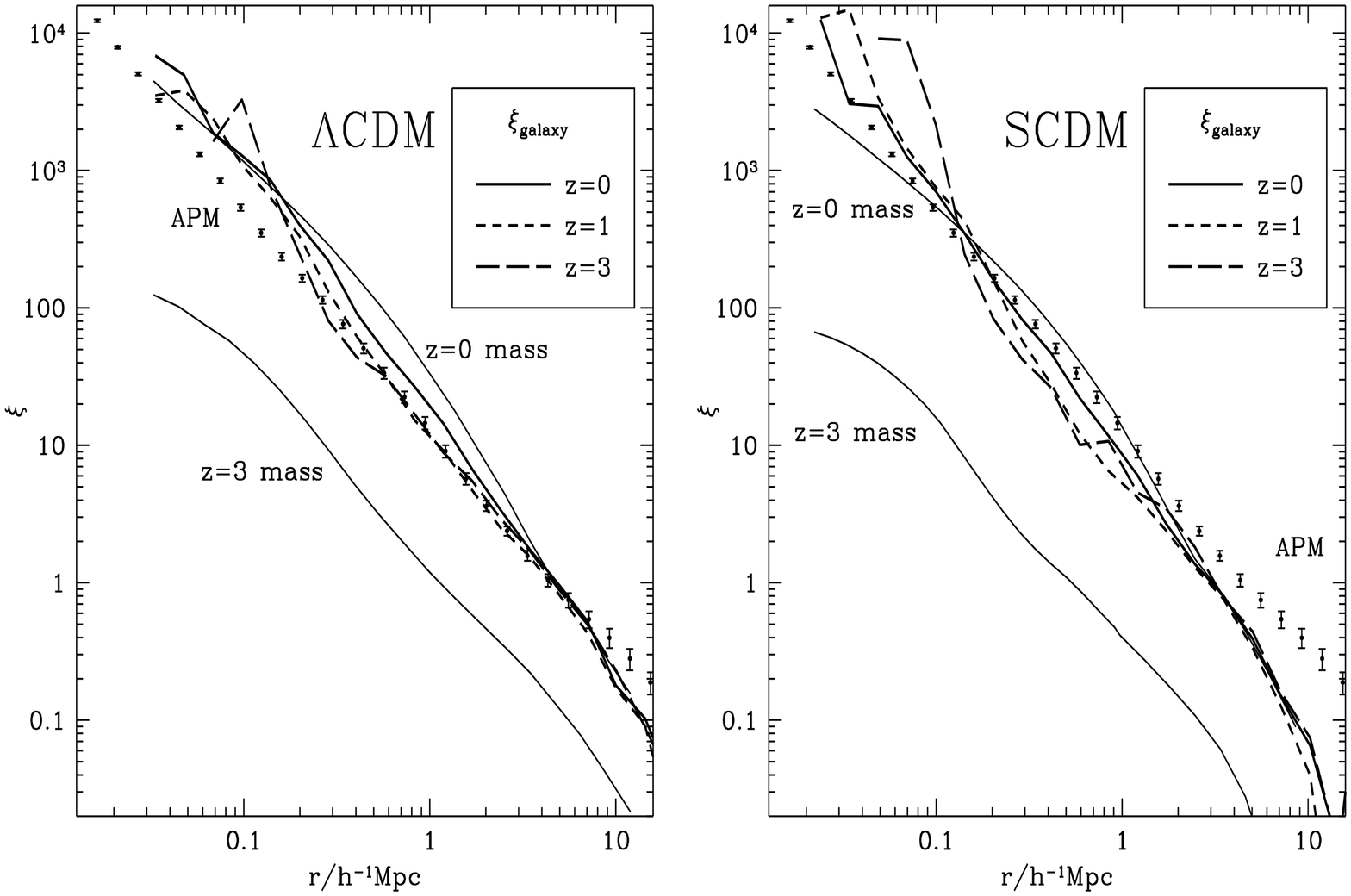,angle=270,height=8.5cm,width=17.5cm}
\caption{Evolution of the galaxy two-point correlation function in the
\LCDM and SCDM models. Also plotted is the observed galaxy correlation
function as determined by Baugh (1996) from deprojecting the angular
correlation function of galaxies in the APM survey. (These data assume no
evolution of the clustering with redshift, as is seen for the simulation
galaxies).  In both models, the galaxy correlation function in comoving
coordinates changes little between redshift 3 and the present, in contrast
to the mass correlation function which changes considerably.  For \LCDM the
galaxies show a significant anti-bias on scales 0.2-3\hmpc and similarly
for SCDM up to $\sim$2\hmpc. See text for more discussion. }
\label{gcorrfn}
\end{figure*}

\section{Clustering properties of the Galaxies}

In this section we examine both the two-point spatial clustering and
the pairwise velocity dispersions of the galaxies as a function of separation
and we compare them to observational determinations.  We also investigate the
dependence of spatial clustering on galaxy mass. We test a virial mass
estimator, that of Heisler, Tremaine, \& Bahcall (1985, hereafter
HTB), on the galaxy 
samples contained within a few of the largest dark matter halos to see how
well the true mass within the virial radius is recovered.

Before proceeding with our analysis, we need to define a suitable galaxy
sample.  As discussed earlier, the galaxies are easily identified  because
of their very high gas density compared to the mean. The exact membership
of the galaxy catalogues obtained using a group finder is relatively
insensitive to the input parameters used to select them.  In this section,
galaxy catalogues for both cosmologies were generated by applying a
``friends-of-friends'' group finder with a linking length of $0.02(1+z)$ on
a subset of gas particles with temperature less than $12000\K$. All gas
particles were counted for the purposes of defining the linking length
itself.  Catalogues were constructed at $z$=0, 1 and 3.  They contain
2263, 2360 and 1502 galaxies respectively for \LCDM and 2089, 1647 and 550
galaxies for SCDM, counting objects of 8 particles or more. 

\subsection{The galaxy autocorrelation function}

Figure 17 shows the redshift evolution of the two-point correlation
function of galaxies in the \LCDM and SCDM models. In both cases, the
galaxy correlation function evolves little over the range $z=$0--3, in
contrast to the mass autocorrelation functions also shown in the figure.
The galaxy correlation functions are essentially unbiased on the largest
scales relative to the $z=0$ mass correlation function but show a distinct
anti-bias on scales below 3\hmpc for \LCDM and 2\hmpc for SCDM, and a
positive bias on scales below $\sim 100$\hkpc in both cases. Also
plotted is Baugh's (1996) measured galaxy autocorrelation function
determined by deprojecting the angular correlation function of galaxies in
the APM catalogue. (The curve plotted assumes that clustering is fixed in
comoving coordinates.)  On small scales, the galaxy correlation function in
the simulations is encouragingly close to the real galaxy correlation
function. On large scales, the SCDM model fails because it lacks sufficient
large-scale power, but the \LCDM model continues to do well.  
The departure of the \LCDM correlation function of both galaxies
and dark matter below the APM result at the largest scales plotted is due
to the small volume of the simulation.  A better comparison between the APM
data and the dark matter correlation function at large separations is shown 
in Fig.~5 of Jenkins \etal (1998). 

To examine the dependence of spatial clustering on galaxy mass, we
subdivided the \LCDM $z=0$ sample into low-, middle- and high-mass
subsamples, each with approximately 750 galaxies.  In terms of the gas
particle number, $N_{part}$, the three subsamples have
$8 \leq N_{part} <50$, $50\leq N_{part} < 95$ and $N_{part}\geq95$ 
respectively, corresponding to $K$-band luminosities in the ranges
($-20$, $-22.3$), ($-22.3$, $-22.8$) and brighter than $-22.8$ .
Figure 18 shows the 2-point correlation function for the three galaxy
subsamples.  The clustering strength shows a clear trend, with the
most massive galaxies having the strongest clustering.  The difference
between the medium- and low-mass samples is much less pronounced than
the difference between the medium- and high-mass samples.

To conclude, we find that the clustering properties of the galaxies in the
simulations are markedly different from those of the dark matter.  In
particular, the galaxy correlation function evolves weakly with redshift
and, coincidentally, appears unbiased on large-scales at the present epoch.
At the present day, the correlation amplitude of the galaxies in both
models is weaker than that of the dark matter at intermediate scales around
1\hmpc and this is particularly pronounced for \LCDM.  The galaxy
correlation function in the
\LCDM model is close to that determined observationally.  
Its shape is rather noisy because of the small sample, but it is much
better fit by a power-law than the mass distribution.  The clustering
strength of galaxies increases with galaxy luminosity, but this effect is
only strong at very bright luminosities.

\begin{figure}
\psfig{file=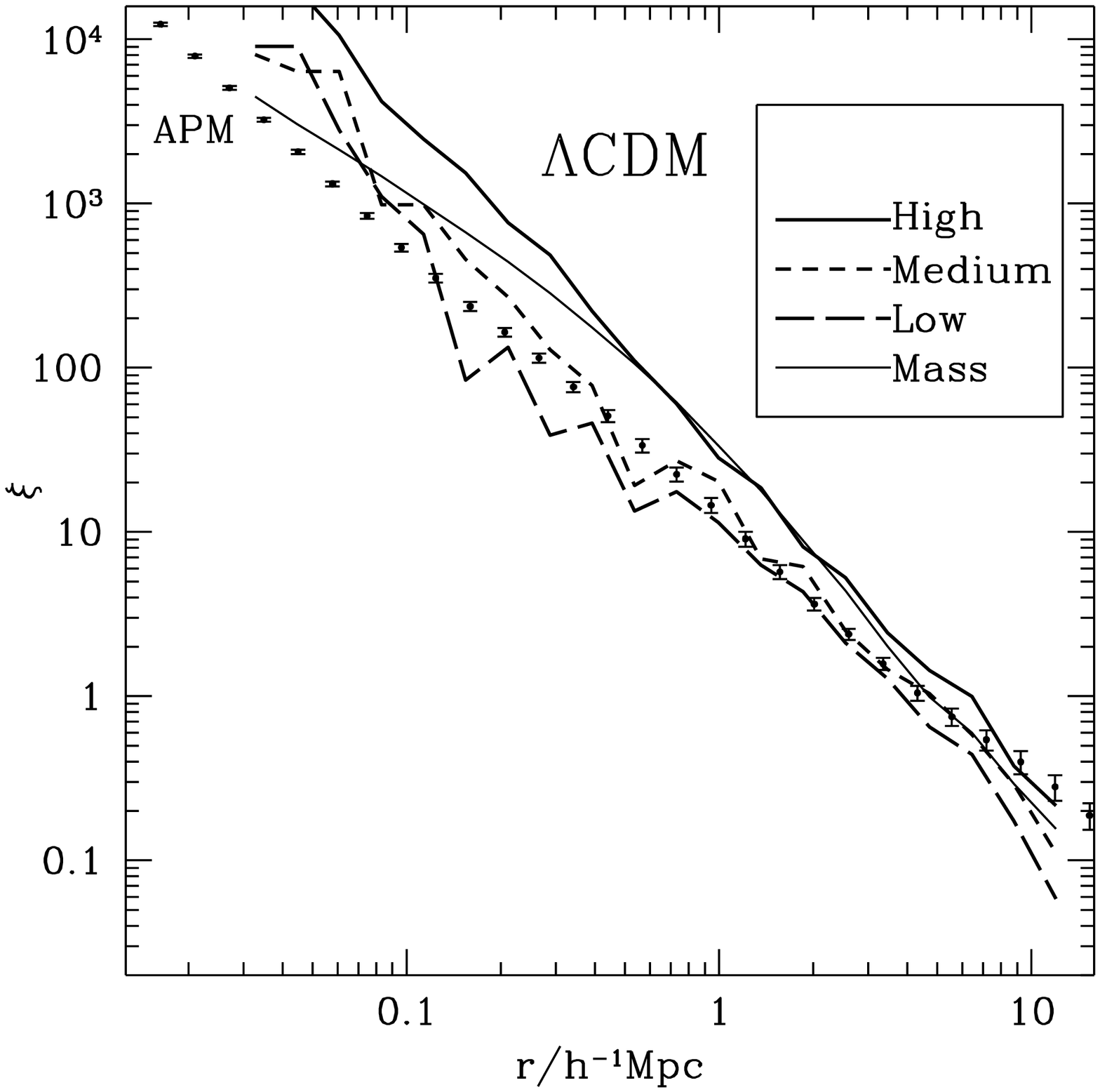,height=8.5cm,width=8.5cm}
\caption{The \LCDM galaxy correlation function for galaxies 
of different mass (or luminosity).  The \LCDM catalogue was sorted by mass
and divided into three approximately equal sample sizes (see text for
details).  The most massive galaxies exhibit stronger clustering than the
low-mass galaxies.
\label{clmasscut}}
\end{figure}

\subsection{The galaxy projected pairwise velocity dispersion.}

In this subsection, we compare the projected pairwise velocity
dispersions of galaxies and dark matter.  As is well-known, a reliable
estimate of this statistic requires a rather large sample volume
(Marzke \etal 1995, Mo, Jing \& Boerner 1997). The sample volumes that
we analyse in this paper are probably too small to provide an accurate
estimate of the global pairwise velocity dispersions, and we will
therefore concentrate on the relative dispersions of galaxies and dark
matter.  Nonetheless, with this caveat in mind, we plot, for interest,
the pairwise velocity dispersions determined from the LCRS catalogue
by Jing, Mo \& Boerner (1998).

Fig.~19 shows the measured projected pairwise velocity dispersions of both
dark matter and galaxies in the \LCDM and SCDM models.  The detailed
definition of the projected pairwise dispersion is given in Jenkins \etal
(1998).  Because of our smaller simulation volumes, we have decreased the
length-scale of the projection from $\pm$25\hmpc to $\pm$10\hmpc and
$\pm$8\hmpc for the \LCDM and SCDM models respectively. We plot the same
LCRS data points (taken from Jing \etal 1998), as plotted in Fig.~11 of
Jenkins \etal (1998).

There is a remarkable difference between the pairwise velocity correlations
of galaxies and those of the dark matter on scales below $\sim2$\hmpc in both
cosmological models. For \LCDM, the galaxy pairwise velocities turn out to
be not much higher than the observational data points.  In fact, the dark
matter pairwise dispersion in our 70\hmpc \LCDM simulation cube is around
100\kms higher than the value obtained by averaging over larger 239.5\hmpc
boxes in Jenkins \etal (1998; Fig.~11). One might infer from this that the
global average for the projected galaxy pairwise dispersion is also
overestimated. The difference between the behaviour of the galaxies and
dark matter becomes smaller with increasing pair separation and is similar
on the largest scales probed by these simulations.

\begin{figure*}
\psfig{file=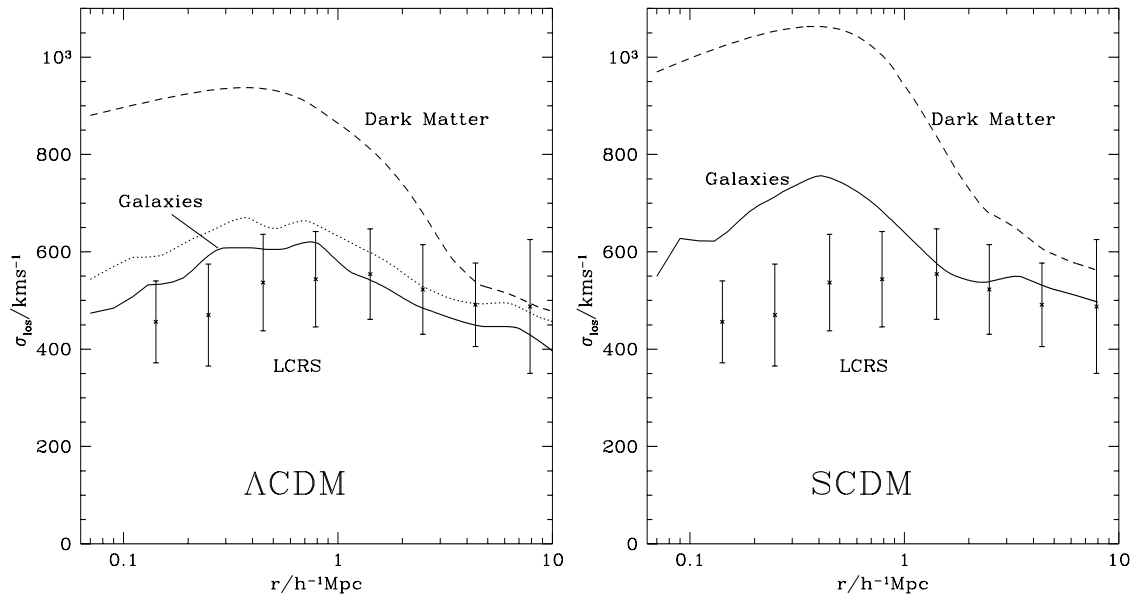,height=8.5cm,width=15cm,angle=270}
\caption{ A comparison of the projected two-point pairwise
velocity dispersion of galaxies and dark matter for \LCDM and SCDM.  The
data points are from Jing \etal (1998) based on the Las Campanas Redshift
Survey.  The galaxy-galaxy pairwise velocities are significantly different
to those of the dark matter.  The dotted line shows the projected two-point
pairwise velocity dispersion of a catalogue made by selecting the nearest
dark matter particle to each galaxy.  The similarity of these dispersions
and those of the galaxies suggests that the differences between galaxies
and dark matter as a whole are mostly due to the way in which galaxies
populate dark halos rather than to an intrinsic velocity bias of galaxies.}
\label{vpcorrfn}
\end{figure*}

Why do the galaxies have a lower pairwise dispersion than the dark matter?
To help answer this question we have constructed a ``shadow galaxy''
catalogue for \LCDM by selecting the nearest dark matter particle to each
galaxy.  The shadow catalogue has, by construction, virtually an identical
2-point correlation function to the galaxies. In Fig.~19 we also plot, as a
dotted line, the projected pairwise dispersion for the shadow catalogue.
This is almost identical to the projected pairwise dispersion of the
galaxies. This shows that the strong difference between the galaxies and
the dark matter arises mostly from the way in which galaxies populate dark
matter halos rather than from a strong intrinsic bias in the velocities of
galaxies relative to the dark matter particles at the same location.  The
fact that the dispersion of the shadow catalogue is everywhere higher than
that of the galaxies indicates that some residual is present, but the
closeness of the curves shows that this is not a large effect. Thus, the
low galaxy velocity dispersion is a statistical effect which may, in
principle, be driven by differences in the ratio of the number of galaxies
to total mass in high- and low-mass halos and also by differences in the
way in which galaxies are distributed within halos compared to the dark
matter.

Our results for \LCDM are very similar to those in Benson \etal (2000a),
but differ in some respects from those of Kauffmann \etal (1999b).  In both
papers, semi-analytical modelling of galaxy formation was used to create
synthetic galaxy catalogues from the same
\LCDM N-body simulation, but the detailed placement of galaxies in
halos was different. 
Generally, in the semi-analytic models the efficiency of galaxy
formation per unit mass of dark matter is a strong function of dark
halo mass and peaks for halos with masses around $10^{12}\Mo$. 
Benson \etal (2000a) found that by populating the dark
matter halos unevenly as their semi-analytical model predicts, not only
does one obtain a galaxy correlation function which matches the APM
result of Baugh (1996) well, but also a comparable difference as 
found here in the projected pairwise dispersion of galaxies 
relative to the dark matter.  Kauffmann \etal (1999b) find a very
similar two-point correlation function, but rather little difference
between the pairwise dispersions of galaxies and dark matter.  As the
analysis of Benson \etal (2000a) shows, this discrepancy can be traced
back to relatively small variations in the precise form of the halo
occupation number predicted in the two models, particularly for halos
with mass $>10^{13}\Mo$.
These generate a large difference in the pairwise dispersions but have a
much weaker effect on other statistics such as the two-point correlation
function.  A detailed comparison between our simulations and
semi-analytical modelling techniques applied to dark matter only versions
of our simulations is given in Benson \etal (2000c).

In summary, we find significant differences in the pairwise velocity
dispersion of galaxies relative to the dark matter on scales below
2\hmpc.  The amplitude of this statistical effect appears sufficient
to reconcile \LCDM with current observations.  The differences in the
velocity dispersions of galaxies and dark matter arise because the
galaxies populate dark halos with a variable efficiency which
decreases for the highest mass halos.

\subsection{Virial theorem mass estimates of Galaxy clusters}

In this section we employ a commonly used virial estimator to see how well
the virial masses can be determined for a few of the most massive dark
halos using the projected positions and velocities of the galaxies located
within the halo.  The estimator we have selected is one of those
discussed by Heisler \etal (1985), specifically,
\begin{equation}
  M_{VT} = {3\pi N\over2G}{\sum_i V_{r,i}^2\over\sum_{i<j}1/R_{ij}},
\end{equation}
where $V_{r,i}$ is the line-of-sight velocity of the $i$th galaxy (in the
observer frame where the cluster is at rest) and $R_{ij}$ is the projected
distance between galaxies $i$ and $j$. HTB show that this estimator is
fairly robust. It does assume, however, that the galaxies trace the dark
matter. If the galaxies are, for instance, more concentrated within a
dark halo than the dark matter then the kinetic energy-like upper sum
in equation (2) will be underestimated, whilst the potential
energy-like sum will be overestimated. The result then is a virial
mass estimate which will be lower than the true mass.  Additional
sources of uncertainty in applying the estimator to galaxy clusters
are that the dark halos are not that well isolated from their
surroundings and may not be in close virial equilibrium particularly
if viewed at an epoch of major mass accretion.

As a sample, we take the 9 most massive clusters from both \LCDM
and SCDM models.  The average number of galaxies in our \LCDM halos is
N=12.4 and N=10.7 for SCDM.  We project each cluster from a large number
of different directions to get the full distribution of the estimated
virial mass. We then average the ratio of the estimated mass to the
true mass for the cluster samples.  Using $\log_{10}(M)$ as the mass
variable we find that the distribution of the estimated mass about the
true mass is very similar in both cosmologies but it is slightly more
peaked in \LCDM than SCDM.  With only nine halos for each model, the
conclusions that we can draw are not very strong.  The distributions
have significant tails towards low masses, but the exact form is
dependent on only one or two of the halos. The estimator shows a bias
in $\log_{10}(M)$ of $0.08\pm0.08$ for \LCDM and $-0.02\pm0.2$ for
SCDM which is consistent with zero.  A more reliable statistic which
does not depend so much on the tail of the distributions is the range
$\pm\Delta \log_{10}M$ about the true mass which includes 3/4 of the
distribution. For \LCDM this is about 0.23 and for SCDM 0.2.  In
Figure 20 we combine the results for \LCDM and SCDM. For the combined
sample, 3/4 of the distribution is included within a range of
$\pm0.21$.  This spread of values is a little larger than the 
corresponding spread found by HTB, who quote a range of 0.2 for N=5,
and 0.15 for N=10 but these are for realisations of virialised Mitchie
models.  This is not perhaps surprising given that the galaxies do not
trace the mass exactly nor are the clusters in complete virial
equilibrium.

\begin{figure}
\psfig{file=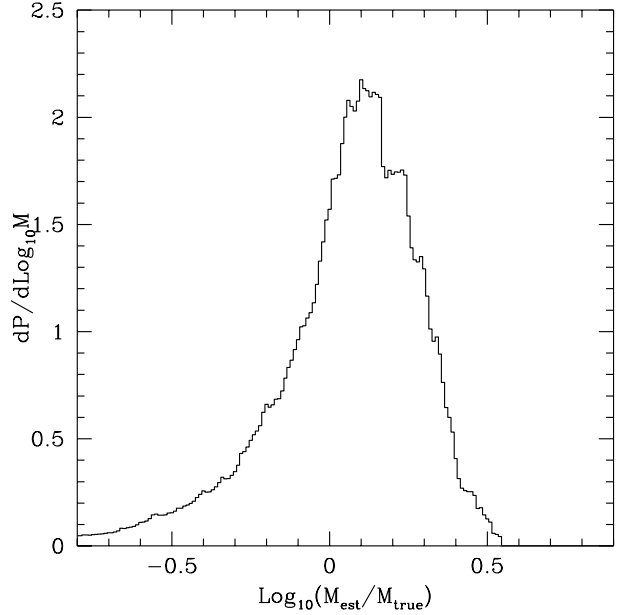,height=8.5cm,width=8.5cm}
\caption{Combined distribution function for the virial mass estimator
(eqn.~2) applied to the galaxies within the largest 9 dark matter halos in
our \LCDM and SCDM simulations. The x-axis shows the difference between the
estimator and the true mass within the virial radius. The estimator is
computed over a large number of randomly chosen directions about each halo.
The average mass of the combined samples is consistent with the true virial
mass; three-quarters of the distribution lies within $\pm0.21$.
\label{dist_vest}}
\end{figure}

\begin{figure}
\psfig{file=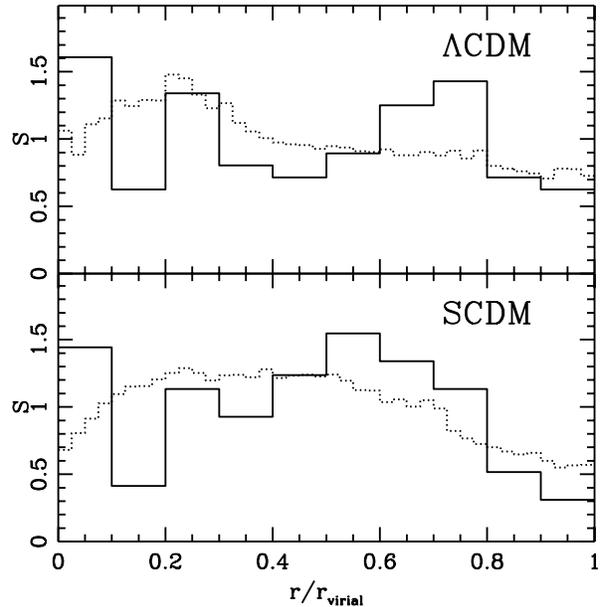,height=8.5cm,width=8.5cm}
\caption{The average radial distribution of galaxies
(solid lines) and mass (dotted lines) within the virial radius of the 9
largest halos from \LCDM and SCDM simulations. All the histograms are
normalised to have unit area underneath.  The distributions are similar.
\label{dist_rad}}
\end{figure}

In Figure 21 we compare the average radial distribution of the
galaxies and mass within the virial radius for the 9 halos from each
cosmology analysed above. The galaxy radial distribution is shown
with a solid line and the mass radial distribution with a dotted
line. There are 112 galaxies contributing to the \LCDM histogram and
97 to SCDM. All the histograms are normalised so as to have unit area
underneath.  Broadly speaking the galaxies and mass have a similar
distribution. A $\chi^2$ analysis of the galaxies and mass
distributions yields $\chi^2\simeq 14$ for 8 d.o.f. for both \LCDM and
SCDM.  Taken individually this is unremarkable with a higher value
occurring approximately 8\% of the time.  The joint $\chi^2$ has a
probability which is only marginally more significant at 3\%.  Without
a larger sample it is difficult to draw any firmer conclusion about
the relative distributions except to note that several hundred
galaxies are needed in order to distinguish the distributions.

In summary, we find that applying the estimator, eqn.~2, to the galaxies in
the largest dark matter halos taken from either of our two simulations, one
can infer the masses of the dark halos in a relatively unbiased way, albeit
with a significant uncertainty for a single determination of a particular
halo. Consistent with this conclusion, the radial distributions of galaxies
and mass are similar. These mass estimates, of course, neglect projection
effects in the identification of cluster galaxies which can add
significantly to the uncertainty (eg van Haarlem, Frenk \& White 1997).  It
is interesting to note that for clusters more massive that $\sim
1.3\times10^{14}$ \Msun, the mean baryon mass fraction in the form of
galaxies is 17\% and 8.4\% for \LCDM and SCDM respectively. These values
differ from the global fractions which are 12\% for \LCDM and 7\% for
SCDM. Thus, the mass-to-light ratios of clusters cannot, in general, be
used to estimate $\Omega_0$ reliably without accurately accounting for the
difference in the efficiency of galaxy formation in clusters compared to
the universe as a whole.

\section{Discussion and conclusions}

Our simulations have many limitations and should be regarded primarily
as illustrative of general trends likely to characterize galaxy
formation in hierarchical CDM models. Most of these limitations stem
from the limited resolution that is attainable with current computing
resources if one wishes to simulate a representative volume of the
universe. Our models do not include a self-consistent treatment of
star formation and feedback. Instead, the ability of gas to cool in
small objects is limited by the resolution of the simulation. This is
clearly a crude approximation. The one virtue of the strategy we have
adopted is that the resolution limit can be adjusted to ensure that
approximately the right amount of gas cools by the end of the
simulation. We have done this for the SCDM simulation and kept the
same resolution for the \LCDM simulation with the result that more gas
cooled in the latter case than is observed in the form of stars in the
real universe. Clearly, further progress will require more realistic,
resolution-independent and physically-based treatments of star
formation and feedback.

Another limitation of our simulations, stemming also from limited
resolution, in this case spatial resolution, is the inadequate
modelling of the internal structure of galaxies.  It seems clear that
a gravitational softening of $10h^{-1}\kpc$ is much too large to
prevent the artificial disruption of galaxies in rich clusters.  Our
strategy of retaining gas that has cooled into galaxies as a gaseous
component rather than turning it into stars as is sometimes done,
partially compensates for this, but it is clear that simulations with
much higher spatial resolution are required. Unlike the problem of
star formation and feedback, this is a limitation that can, in
principle, be overcome by increased computing power. For example, a
simulation similar to ours but with a gravitational softening of only
$2h^{-1}\kpc$ would require $\sim 100,000$ timesteps at least for the
particles in the densest regions.

In spite of these limitations, our simulations demonstrate the
potential of direct simulation for realistic modelling of galaxy
formation. This study is the first to follow the formation and
evolution of a large number of galaxies in a volume, a cube of side
$100\Mpc$, which may be regarded as representative. Over 2000 galaxies
brighter than $L_*$ formed in each of our simulations.  Both in this
paper and elsewhere (Pearce \etal 1999) we have shown that the
galaxies that form in the simulations have a spatial distribution that
is consistent with observations not only locally, but also at high
redshift.

Here, we have used our simulations to illustrate the hierarchical
build up of galaxies and to give an indication of the rate at which
gas is expected to cool into fragments which subsequently merge to
form larger galaxies.  The aggregate star formation rate in the
fragments that end up in a massive galaxy today adds up to a large
value and is sharply peaked at high redshift.  By contrast, the star
formation rate of a typical galaxy today is more modest and spread out
in time. The mean star formation rate in the simulation as a whole
rises at early times, reaches a broad maximum between $z=2$ and $z=1$
and declines towards the present, in a manner reminiscent of the data
of Steidel \etal (1999). Although this behaviour is, to a large
extent, determined by resolution effects, the fact that it resembles
the data suggests that the bulk properties of the model galaxy
population today are not too unrealistic.  In the \LCDM simulation, the
first substantial galaxies form at $z\simeq 5$. The galaxy population
builds up rapidly until $z=1$ and there is a marked decline in the
rate of change of the galaxy mass function after $z=0.5$. In the SCDM
simulation there is more evolution at recent times.  All these trends
have been previously emphasized in semi-analytic models of galaxy
formation (e.g. Cole \etal 1994, Kauffmann 1996).  At the present
day both simulations match the bright end of the observed K-band
luminosity function for suitable values of the mass-to-light ratio.

The galaxy autocorrelation functions evolve little with redshift over
the range $z=0-3$.  The galaxy correlation function agrees closely
with that of the dark matter at separations of $10h^{-1}\Mpc$ at $z=0$
but differs significantly on smaller scales.  For both, the \LCDM and SCDM 
models, the
galaxy correlation function appears closer to a power law than the
mass correlation function.  The amplitude of the galaxy correlation
function increases with the mass of the galaxies.  The projected
pairwise velocity distribution of the galaxies is significantly lower,
particularly in \LCDM, than that of the dark matter.

Within the most massive dark matter halos the galaxies trace the dark
matter faithfully.  A virial mass estimator applied to the clusters
correctly infers the appropriate dark matter mass albeit with a large
dispersion for a single determination.  However, galaxy formation is
more efficient in such clusters than in the simulation
as a whole, suggesting that the cosmic density parameter cannot safely
be inferred from the mass-to-light ratios of clusters and the mean
luminosity density of the Universe.

\section*{Acknowledgements}

We would like to thank Shaun Cole, Eric Tittley, Scott Kay and Andrew
Benson for providing analysis software and data.  The work presented in
this paper was carried out as part of the programme of the Virgo
Supercomputing Consortium (http://star-www.dur.ac.uk/~frazerp/virgo/)
using computers based at the Computing Centre of the Max-Planck Society in
Garching and at the Edinburgh Parallel Computing Centre.  This work was
supported in part by grants from PPARC, EPSRC, and the EC TMR network for
``Galaxy Formation and Evolution''.  CSF acknowldeges a Leverhulme Research
Fellowship and PAT holds a PPARC Lecturer Fellowship. HMPC is supported by
NSERC of Canada.

\label{lastpage}

\begin{thebibliography}{}

\bibitem[Abel Anninos Norman \& Zhang 1998]{1998ApJ...508..518A} Abel, T. , 
Anninos, P. , Norman, M. L., Zhang, Y.  1998, \apj, 508, 518 

\bibitem[Baugh 96]{Baugh} Baugh, C. M., 1996, \MN, 280, 267

\bibitem[Baugh Cole Frenk \& Lacey 1998]{BCFL} Baugh, C. M., 
Cole, S., Frenk, C. S., Lacey, C. G. 1998, \ApJ, 498, 504 

\bibitem[5]{5} Benson, A. J., Baugh, C. M., Cole, S., Frenk, C. S., Lacey,
C., 2000a, \MN, 316, 107

\bibitem[4]{4} Benson, A. J., Cole, S., Frenk, C. S., Baugh, C. M., Lacey,
C., 2000b, \MN, 311, 793

\bibitem[Benson \etal 2000a]{Be2000} Benson, A. J., Pearce, F. R.,
Frenk, C. S., Baugh, C.,  Jenkins, A., 2000c,
\MN, in press, astro-ph/9912220 

\bibitem[6]{6} Blanton, M., Cen, R., Ostriker, J. P., Strauss, M. A., 1999,
\ApJ, 522, 590

\bibitem[Bruzual,  Charlot 1993]{BC93} Bruzual, G., Charlot, S., 1993, \ApJ, 405, 538 

\bibitem[Carlberg Couchman,  Thomas 1990]{1990ApJ...352L..29C} Carlberg, 
R. G., Couchman, H. M. P.,  Thomas, P. A. 1990, \apjl, 352, L29 

\bibitem[Cen,  Ostriker 1996]{Cen}
Cen, R., Ostriker, J., 1996, \ApJ, 464, 270

\bibitem[c91]{c91} Cole, S., 1991, \ApJ, 367, 45

\bibitem[Cole \etal 2000]{Cole2000}
Cole, S. \etal, 2000, \MN, in press

\bibitem[7]{7} Copi, C. J., Schramm, D. N., Turner, M. S., 1995, \ApJ, 455,
95

\bibitem[8]{8} Couchman, H. M. P., Thomas, P. A., Pearce, F. R., 1995,
\ApJ, 452, 797

\bibitem[Croft]{Croft} Croft, R. A. C., Di Matteo, T, Dave\'e, R.,
Hernquist, L., Katz, N., Fardal, M. A., Weinberg, D. H., submitted \ApJ,
astro-ph/0010345 

\bibitem[9]{9} Davis, M., Efstathiou, G., Frenk, C. S., White, S. D. M., 1985, \ApJ,
292, 371

\bibitem[Diaferio et al. <1999>]{diaferio99}Diaferio~A., Kauffmann~G.,
Colberg~J.~M., White~S.~D.~M., 1999, \MN, 307, 537 

\bibitem[Eke, Cole,  Frenk 1996]{Eke96}
Eke, V. R., Cole, S., Frenk, C. S., 1996, \MN, 282, 263

\bibitem[Evrard Summers,  Davis 1994]{ESD} Evrard, A. E., 
Summers, F. J.,  Davis, M.  1994, \ApJ, 422, 11 

\bibitem[Evrard Metzler,  Navarro 1996]{1996ApJ...469..494E} Evrard, A. 
E., Metzler, C. A.,  Navarro, J. F. 1996, \apj, 469, 494 

\bibitem[Frenk \etal 1996]{FEWS}
Frenk, C. S., Evrard, A. E., White, S. D. M., Summers, F. J., 1996,
\ApJ, 472, 460

\bibitem[Fukugita, Hogan and Peebles (1998)]{1998ApJ...503..518F} Fukugita, 
M., Hogan, C. J., Peebles, P. J. E. 1998, \ApJ, 503, 518 

\bibitem[Gardner \etal 1997]{gsfc97} Gardner, J. P., Sharples, R. M.,
Frenk, C. S., Carrasco, E., 1997, \ApJ, 480, 99

\bibitem[Gerritsen,  Icke 1997]{1997A&A...325..972G} Gerritsen,
J. P. E.,  Icke, V. 1997, \aap, 325, 972 

\bibitem[Gingold,  Monaghan 1977]{Gingold} Gingold, R. A.,  
Monaghan, J. J. 1977, \MN, 181, 375 

\bibitem[Guiderdoni Hivon Bouchet,  Maffei 1998]{1998MNRAS.295..877G} 
Guiderdoni, B. , Hivon, E. , Bouchet, F. R.,  Maffei, B.  1998, \mnras, 
295, 877

\bibitem[van Haarlem, Frenk \& White 1996]{vh96} van Haarlem, M.,
Frenk, C.S., White, S. D. M., 287 \MN, 817.

\bibitem[Heisler, Tremaine,  Bahcall 1985]{HTB}Heisler, J., Tremaine, S.,  Bahcall, J.,
1985,\ApJ, 298, 8

\bibitem[JMB98]{Jing98}Jing, Y. P., Mo, H. J.,  Boerner, G., 1998, \ApJ, 494, 1

\bibitem[Jenkins et al. 1998]{Jenkins97} Jenkins, A., \etal 
1998, \ApJ, 499, 20 

\bibitem[Jenkins et al. 2000]{Jenkins2000} Jenkins, A., \etal 
2000, in press \MN, astro-ph/0005260

\bibitem[Katz 1992]{Katz92}
Katz, N., 1992, \ApJ, 391, 502

\bibitem[Katz, Hernquist,  Weinberg 1992]{KHW}
Katz, N., Hernquist, L., Weinberg, D. H., 1992, \ApJ, 399, L109

\bibitem[Katz, Hernquist and Weinberg (1999)]{1999ApJ...523..463K} Katz, 
N., Hernquist, L., Weinberg, D. H., 1999, \apj, 523, 463 

\bibitem[14b]{14b} Katz, N., Weinberg, D. H.,  Hernquist, L., 1996, \ApJS,
105, 19 

\bibitem[Kay \etal]{Kay}Kay, S. T., Pearce, F. R., Jenkins, A., Frenk, C. S., 
White, S. D. M., Thomas, P. A., Couchman, H. M. P., 2000, \MN, 316, 374

\bibitem[13z]{13z} Kauffmann, G. 1996, \MN, 281, 475. 

\bibitem[13a]{13a} Kauffmann, G., Colberg, J., Diaferio, A., White,
S. D. M., 1999a, \MN, 303, 188

\bibitem[13b]{13b} Kauffmann, G., Colberg, J., Diaferio, A., White,
S. D. M.,
1999b, \MN, 307, 529 

\bibitem[13c]{13c} Kauffmann, G., Nusser A., Steinmetz M., 1997, \MN, 286, 795

\bibitem[Lucy 1977]{Lucy} Lucy, L. B. 1977, \AJ, 82, 1013

\bibitem[Marzke95]{Marzke95} Marzke, R. O., Geller, M. J., daCosta, L. N., 
Huchra, J. P., 1995, \AJ, 110,477

\bibitem[Metzler,  Evrard 1994]{1994ApJ...437..564M} Metzler, C. A.,  
Evrard, A. E. 1994, \apj, 437, 564 

\bibitem[Mihos,  Hernquist 1994]{1994ApJ...437..611M} Mihos, C. J. ,  
Hernquist, L.  1994, \apj, 437, 611 

\bibitem[Mihos,  Hernquist 1996]{Mihos96}
Mihos, C. J.,,  Hernquist, L., 1996, \ApJ, 464, 641

\bibitem[MJB97]{MJB97} Mo, H. J., Jing, Y. P.,  Boerner, G., 1997, \MN, 286, 979

\bibitem[Navarro,  Steinmetz 1997]{NS97}
Navarro, J. F., Steinmetz, M., 1997, \ApJ, 478, 13

\bibitem[Navarro,  White 1993]{Navarro93}
Navarro, J. F., White, S. D. M., 1993, \MN, 265, 271

\bibitem[Pearce,  Couchman 1997]{PC97}
Pearce, F. R., Couchman, H. M. P., 1997, New Astronomy, 2, 411

\bibitem[Pearce \etal 1999]{Pe1999} Pearce, F. R. \etal, 1999, \ApJL, 521, 99

\bibitem[Pearce, Thomas, Couchman and Edge (2000)]{2000MNRAS.317.1029P} 
Pearce, F. R., Thomas, P. A., Couchman, H. M. P.,  Edge, A. C. 
2000, \MN, 317, 1029

\bibitem[Press,  Schechter 1974]{Press}
Press, W. H., Schechter, P., 1974, \ApJ, 187, 425

\bibitem[RT]{RT} Ritchie, B. W., Thomas, P. A., 2000, submitted \MN, 
astroph/0005357

\bibitem[Somerville \etal]{Somerville} Somerville, R. S., Primack, J. 
1999, \MN, 310, 1087. 

\bibitem[Steidel et al.\ (1999)]{1999ApJ...519....1S} Steidel, C. C., 
Adelberger, K. L., Giavalisco, M., Dickinson, M., Pettini, M. 1999, 
\apj, 519, 1 

\bibitem[SM95]{SM95}  Steinmetz, M., M\"uller, E., 1995, \MN, 276, 549 

\bibitem[Thacker \etal 2000]{Thacker00}Thacker, R. J., Tittley, E. R.,
Pearce, F. R., Couchman, H. M. P., Thomas, P. A., 2000, \MN, in press,
astro-ph/9809221

\bibitem[Tytler, O'Meara, Suzuki and Lubin (2000)]{2000PhST...85...12T} 
Tytler, D., O'Meara, J. M., Suzuki, N.,  Lubin, D. 2000, Physica 
Scripta Volume T, 85, 12 

\bibitem[Weil Eke,  Efstathiou 1998]{1998MNRAS.300..773W} Weil, M. L., 
Eke, V. R.,  Efstathiou, G. 1998, \mnras, 300, 773 

\bibitem[White,  Frenk 1991]{White91}
White, S. D. M., Frenk, C. S., 1991, \ApJ, 379, 52

\bibitem[wr]{wr} White, S. D. M., Rees, M. J., 1978, \MN, 183, 341

\end{thebibliography}
\end{document}